\documentstyle[12pt,epsf]{article}
\newcommand{\la}[1]{\label{#1}}

\newcommand{\pp}[1]{\langle\phi^\dagger\phi(#1)\rangle}

\newlength{\numlen}

\newcommand{\cen}[1]{\multicolumn{1}{|c|}{#1}}
 
\newcommand{\figysize}{14.5cm}
\newcommand{\figtopspace}{\vspace*{-1.3cm}}
\newcommand{\figbottomspace}{\vspace*{-5.5cm}}

\newlength{\indexlength}

\newcommand{\be}{\begin{equation}}
\newcommand{\ee}{\end{equation}}
\newcommand{\ba}{\begin{eqnarray}}
\newcommand{\ea}{\end{eqnarray}}

\newcommand{\rmi}[1]{{\mbox{\scriptsize #1}}}
\newcommand{\pdp}{\langle\phi^\dagger\phi\rangle}

\newcommand{\etal}{{et al.\ }}
\newcommand{\eq}{eq.~}

\newcommand{\fig}{Fig.~}
\newcommand{\figs}{Figs.~}

\newcommand{\nr}[1]{(\ref{#1})}

\newcommand{\tr}{{\rm Tr\,}}

\newcommand{\half}{{\scriptstyle{1\over2}}}

\newcommand{\fr}[2]{{\frac{#1}{#2}}}

\newcommand{\msbar}{\overline{\mbox{\rm MS}}}

\def\lsi{\raise0.3ex\hbox{$<$\kern-0.75em\raise-1.1ex\hbox{$\sim$}}}
\def\gsi{\raise0.3ex\hbox{$>$\kern-0.75em\raise-1.1ex\hbox{$\sim$}}}
\newcommand{\lsim}{\mathop{\lsi}}
\newcommand{\gsim}{\mathop{\gsi}}

\makeatletter \@addtoreset{equation}{section} \makeatother

\begin{document}
\begin{titlepage}
\begin{flushright}
BI-TP 96/54\\
CERN-TH/96-334\\
HD-THEP-96-48\\
hep-lat/9612006\\
December 6, 1996\\
\end{flushright}
\begin{centering}
\vfill

{\bf A NON-PERTURBATIVE ANALYSIS OF THE FINITE $T$ PHASE TRANSITION
IN SU(2)$\times$U(1) ELECTROWEAK THEORY}
\vspace{0.8cm}

K. Kajantie$^{\rm a,b}$\footnote{keijo.kajantie@cern.ch},
M. Laine$^{\rm c}$\footnote{m.laine@thphys.uni-heidelberg.de},
K. Rummukainen$^{\rm d}$\footnote{kari@physik.uni-bielefeld.de} and
M. Shaposhnikov$^{\rm a}$\footnote{mshaposh@nxth04.cern.ch} \\

\vspace{0.3cm}
{\em $^{\rm a}$Theory Division, CERN, CH-1211 Geneva 23,
Switzerland\\}
\vspace{0.3cm}
{\em $^{\rm b}$Department of Physics,
P.O.Box 9, 00014 University of Helsinki, Finland\\}
\vspace{0.3cm}
{\em $^{\rm c}$Institut f\"ur Theoretische Physik,
Philosophenweg 16,\\
D-69120 Heidelberg, Germany\\}
\vspace{0.3cm}
{\em $^{\rm d}$Fakult\"at f\"ur Physik, Postfach 100131, D-33501
Bielefeld,
Germany}

\vspace{0.7cm}
{\bf Abstract}

\end{centering}

\vspace{0.3cm}\noindent
The continuum 3d SU(2)$\times$U(1)+Higgs theory is an
effective theory for a large class of 4d high-temperature gauge
theories, including the minimal standard model and some of its
supersymmetric extensions. We study the effects
of the U(1) subgroup using lattice Monte Carlo techniques.
When $g'^2/g^2$ is increased from the zero corresponding
to pure SU(2)+Higgs theory, the phase transition gets stronger.
However, the increase in the strength is close to what is expected
perturbatively, and the qualitative features of the phase
diagram remain the same as for $g'^2=0$. In particular,
the first order transition still disappears for $m_H>m_{H,c}$.
We measure the photon mass and mixing angle, and find that the
mass vanishes in both phases within the statistical errors.

\vfill
\noindent

\end{titlepage}

\section{Introduction}
For reliable computations in the cosmology of the very early
universe, it is indispensable to know the equation of state of the
matter governed by the laws of the one correct -- so far unknown --
physical electroweak theory. For a large class of candidate theories
an effective theory, the 3d SU(2)$\times$U(1) + fundamental Higgs
theory, can be derived by analytic perturbative computations
\cite{generic,mssm}. This effective theory gives an approximate 
but in many cases
accurate description of the electroweak phase transition. The
non-perturbative effects always encountered in a finite $T$ context
are isolated in the effective theory, which has to be solved
numerically. Up to now, all non-perturbative efforts [3--8]
were concentrated on an
even simpler theory, without the U(1) factor.
The 4d lattice simulations of the bosonic sector
of the standard model \cite{desylattice1}
also do not take into account the U(1) group. (The present status of
lattice Monte Carlo studies of the SU(2)+Higgs model both in the
finite temperature 4d theory and in the 3d effective theory is
reviewed in \cite{karilat96}.) The purpose of this article is to
complete the study by including the U(1) subgroup.

At first sight, the inclusion of the U(1) factor is
harmless and may be done perturbatively. However, 
this need not be so in the vicinity
of the phase transition. The dimensionless high-temperature expansion
parameter associated with the U(1) group is $\sim g'^2 T/m$, where
$g'$ is the U(1) gauge coupling and $m$ is a typical mass of a (hyper)
charged particle in 3d. This parameter is indeed small deep in the
broken phase (where $m$ is the W boson mass, $m\sim gv$) 
and high in the symmetric phase (where $m$ is the mass of a 
hypercharged scalar excitation, $m \sim g T$). 
However, near the critical temperature the scalar mass may
be small and perturbation theory breaks down.
Thus a Monte Carlo study is needed.

The inclusion of the U(1) factor also changes qualitatively the 
mass spectrum of the SU(2)+Higgs theory, 
as there is now an excitation (the photon) which is
perturbatively massless in both phases. There are several
interesting questions related to this excitation, for instance
whether there might be non-perturbative mass generation for it
and how the Weinberg mixing angle (defined in 
a suitable gauge independent manner) behaves in the vicinity of
the phase transition.  

It is important to keep in mind the separation of the two stages in
the effective field theory approach. The first stage establishes the
values of the parameters of the 3d effective theory in terms of the
physical 4d parameters by means of a perturbative computation. This has
already been done for the minimal standard model (MSM) 
in \cite{generic} and for the MSSM in
\cite{mssm}. The second stage is a non-perturbative lattice study of
the 3d theory as such, performed in the present paper.

Since the simulations are rather demanding in computer time, we do
not chart the entire critical surface but perform simulations only at
two new points of the parameter space. The nonzero value of the 
U(1) gauge coupling $g_3'^2=0.3g_3^2$
comes from the ratio of the physical values of $m_W$, $m_Z$
in the minimal standard model. For the scalar
self-coupling we choose two values:
\begin{itemize}
\item $\lambda_3=0.06444g_3^2$, which in the notation of~\cite{nonpert}
corresponds to $m_H^* = 60$ GeV (for the actual physical Higgs pole mass
$m_H$ in different 4d theories, see Sec.~\ref{4d3d}). 
This value of the ``Higgs mass'' was the
most precisely studied case in the SU(2)+Higgs model. The first order 
nature of the phase transition here is clear. 
\item $\lambda_3=0.62402g_3^2$, corresponding to $m_H^*=180$ GeV. In the 
SU(2)+Higgs model first or second order phase transitions are absent
then~\cite{isthere?}. We find
that the same statement is true when the U(1) interactions are added.
\end{itemize}

The paper is organized as follows. Sec.~2 is the formulation
of the 3d problem in the continuum, Sec.~3 gives examples of
4d$\to$3d connections and in Sec.~4 we describe how
the theory is discretized.  
Some perturbative results are in Sec.~5. 
Sec.~6 contains the lattice results and a comparison 
with perturbation theory, and the conclusions are in Sec.~7.

\section{Formulation of the problem in 3d}

The Lagrangian of the effective theory (we keep only
super-renormalizable interactions; the accuracy of this approximation
in the MSM is discussed in \cite{generic,jkp}) is:
\be
L={1\over4} F_{ij}^aF_{ij}^a+{1\over4} B_{ij}B_{ij}+
(D_i\phi)^\dagger D_i\phi+m_3^2\phi^\dagger\phi+
\lambda_3(\phi^\dagger\phi)^2,
\label{contaction}
\ee
where
\ba
F_{ij}^a&=&\partial_iA_j^a-\partial_jA_i^a-g_3
\epsilon^{abc}A_i^bA_j^c
\nonumber\\
B_{ij}&=&\partial_iB_j-\partial_jB_i
\nonumber\\
D_i&=&\partial_i+ig_3A_i+ig'_3B_i/2,\qquad
A_i={1\over2}\sigma_aA_i^a,
\nonumber\\
\phi&=&\left(\matrix{\phi_1\cr\phi_2\cr}\right)\equiv
\left(\matrix{\phi^+\cr\phi^0\cr}\right).
\ea
The SU(2)$\times$U(1) local gauge transformation is
\be
\phi(x)\to e^{i\alpha(x)}G(x)\phi(x).
\la{phitransf}
\ee
The four parameters of the theory,  $g_3^2$, $g_3'^2$, $\lambda_3$
and $m_3^2$ are definite computable functions of the underlying 4d
parameters and the temperature.

{}From the 3d point of view, the coupling constants $g_3^2$, $g_3'^2$ and
$\lambda_3$ are renormalization group invariant, whereas the mass
parameter $m_3^2$ gets renormalized at 2-loop level. In the $\msbar$
scheme, the renormalized part of $m_3^2$ is of the form
\be
m_3^2(\mu)=  {f_{2m}\over16\pi^2}\log\frac{\Lambda_m}{\mu},
\la{m3}
\ee
where
\be
f_{2m} =
{51\over16}g_3^4
-\frac{9}{8}g_3^2g_3'^2-\frac{5}{16}g_3'^4
+9\lambda_3 g_3^2
+3\lambda_3 g_3'^2-12\lambda_3^2,
\la{f2m}
\ee
and $\Lambda_m$ is a constant specifying the theory.

All the four parameters of the 3d SU(2)$\times$U(1)+Higgs theory are
dimensionful. One may measure all the dimensionful observables in
terms of one of the coupling constants, say $g_3^2$. The dynamics
then depends on the three dimensionless parameters $x$, $y$ and $z$,
defined as
\be
x\equiv {\lambda_3\over g_3^2},\qquad
y\equiv {m_3^2(g_3^2)\over g_3^4},\qquad
z\equiv {g_3'^2\over g_3^2}.
\la{3dvariables}
\ee
These quantities are renormalization group invariant.

The question we are going to address is: "How does the phase
structure and how do the gauge invariant operator expectation values
depend on $z \ge 0$?" When $z=0$ the U(1) interactions are decoupled
and we get an SU(2)+Higgs theory, the phase structure of which
has been studied in \cite{nonpert,isthere?}. 
In the opposite limit, $z\rightarrow
\infty$ with $x/z$ fixed, we get a 3d U(1)+Higgs model with global SU(2) 
symmetry. The physical value of $z$ lies in between and is
related to the Weinberg angle, $z\approx m_Z^2/m_W^2-1\approx 0.3$.

The main characteristic of the 3d SU(2)+Higgs theory is the existence
of a critical line $y = y_c(x)$ for $0 <x \le x_c \approx 1/8$, along
which the system has a first order transition. The transition gets
weaker with increasing $x$ and eventually terminates in a second
order transition possibly of the Ising universality class at $x_c$.
At $x > x_c$ there is no transition or a transition of higher than
second order. If there is no transition,
the two phases are analytically connected, which
is possible since there is no local gauge invariant order parameter which
would vanish in either of them [12--15].
The signal of
the first order transition is that along $y = y_c(x)$ the system can
coexist in two phases with different values of various gauge
invariant operators like $\pdp$.

Correspondingly, for the SU(2)$\times$U(1)+Higgs theory one expects
that there is a critical surface $y=y_c(x,z)$, which for $z=0$
coincides with the previous curve $y=y_c(x)$. For larger $x$ the surface
should terminate in a line of second order transitions, $x=x_c(z),
x_c(0)\approx1/8$ (Fig.1). Indeed, the statement about the
non-existence of a local gauge invariant order parameter is unchanged. This
expectation has to be confirmed numerically, and the precise
magnitude of the effects of making $z>0$ on various physical
quantities have to be determined.
Expressing everything in dimensionless form by scaling by powers
of $g_3^2$, the quantities we are interested in are
\begin{itemize}
\item The critical curve $y=y_c(x,z)$,
\item The jump $\Delta\ell_3$ of the order parameter like quantity
$\ell_3\equiv\langle\phi^\dagger\phi(g_3^2)\rangle/g_3^2$ between
the broken and symmetric phases at $y_c$. In perturbation
theory,
$\Delta\ell_3\sim\half\phi^2_b(y_c)/g_3^2$, where $\phi_b$ is
the location of the broken minimum in, say, the Landau gauge,
\item The interface tension $\sigma_3$, defined in perturbation
theory by
\be
\sigma_3=\int_0^{\phi_b/g_3}\,d(\phi/g_3)
\sqrt{2V(\phi/g_3)/g_3^6}.
\ee
\end{itemize}

\begin{figure}[t]
\vspace*{1.0cm}
\hspace{1cm}
\epsfysize=8.5cm
\centerline{\epsffile{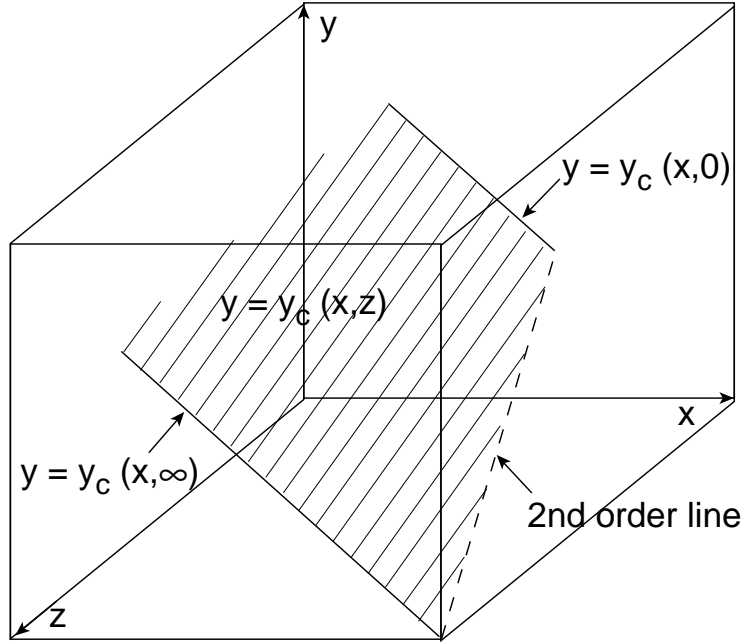}}
\caption[a]{The expected schematic phase diagram of
the SU(2)$\times$U(1)+Higgs theory in the space of the parameters of
eq.(\ref{3dvariables}). There is a 1st order 
transition which terminates in a line of 2nd order transitions.}
\la{cond5}
\end{figure}

A non-perturbative study of the dynamics of the theory is
based on a study of the expectation values and correlators
of low-dimensional gauge invariant operators, so for
completeness we will review these here.
It is convenient to introduce a matrix
parametrisation of the Higgs field by writing
\be
\Phi=\biggl((\tilde\phi)(\phi)\biggr)\equiv
\left(\matrix{\phi_2^*&\phi_1\cr-\phi_1^*&\phi_2\cr}\right).
\la{phimatrix2}
\ee
Under an SU(2)$\times$U(1) gauge transformation $\Phi$
transforms according to
\be
\Phi(x)\to G(x)\Phi e^{-i\alpha(x)\sigma_3}.
\label{gt}
\ee
The global transformation
\be
\Phi(x)\to \Phi(x)C^{-1}
\la{custodial}
\ee
where $C$ is an SU(2) matrix, 
will be seen to be an isospin custodial rotation of the vector
states $W_i^a$.

In order of increasing dimensionality the list of gauge-invariant
operators is:

\begin{itemize}
\item Dim = 1: $\phi^\dagger\phi$. Note that
$\tilde\phi^\dagger\phi=0$;
there is only one singlet.
\item Dim = 3/2: $B_{ij}$, the U(1) field.
\item Dim = 2: there are two independent gauge invariant operators,
namely a vector $\phi^\dagger D_i\phi-(D_i\phi)^\dagger\phi$ and a
scalar $(\phi^\dagger\phi)^2$. For the SU(2)+Higgs case ($z=0$)
there are two extra vector operators. In terms of the matrix
representation
(\ref{phimatrix2}) the vector operators invariant under the SU(2)
subgroup are
\ba
W_i^a&=&i\tr \Phi^\dagger D_i\Phi\sigma_a
\nonumber\\
&=&\fr12 i \tr [\Phi^\dagger D_i\Phi-(D_i\Phi)^\dagger\Phi]\sigma_a.
\la{wia}
\ea
Under the U(1) gauge transformation (\ref{gt}) (which is a rotation
around the 3-axis in isospace)
\be
W_i^a\to i\tr\Phi^\dagger D_i\Phi(e^{-i\theta\sigma_3}\sigma_a
e^{i\theta\sigma_3}),
\ee
so that $W_i^{1,2}$ are not gauge-invariant but  $W_i^3$ is.
Note that
\be
\tr [\Phi^\dagger D_i\Phi+(D_i\Phi)^\dagger\Phi]=2 \partial_i
(\phi^\dagger\phi)
\ee
and does not bring in anything new.
We remark also that in the SU(2)$\times$U(1) theory the custodial
symmetry is explicitly broken.  Under the custodial
transformation (\ref{custodial}) the operators in (\ref{wia}) transform
as vectors in the space of $a=1,2,3$,
\be
W_i^a\to i\tr \Phi^\dagger D_i\Phi C^{-1}\sigma_a C,
\ee
and thus a mixing of gauge-invariant and gauge-noninvariant
operators occurs.

\item Dim =  5/2: $\phi^\dagger F_{ij}\phi$, $\phi^\dagger\phi
B_{ij}$.
\item Dim = 3: Terms in the action.
\item Dim = 7/2: The parity violating $J^{PC}=0^{-+}$ terms
\ba
P_1&=& i\epsilon_{ijk}[\phi^\dagger D_k\phi-(D_k\phi)^\dagger\phi]
B_{ij}, \la{parityop1} \\
P_2&=& i\epsilon_{ijk}[\phi^\dagger F_{ij}D_k\phi-(D_k\phi)^\dagger
F_{ij}\phi]. \la{parityop2}
\ea
The list of parity odd operators can be extended by two non-local
operators,
\be
e^{i2\pi N_\rmi{CS}},
\label{csop}
\ee
where
\be
N_\rmi{CS2}={g^2\over16\pi^2}\int d^3x\,
\epsilon_{ijk}\tr\biggl(A_iF_{jk}-\fr23 igA_iA_jA_k\biggr)
\la{cs}
\ee
and
\be
N_\rmi{CS1}={g'^2\over16\pi^2}\int d^3x\,
\epsilon_{ijk}B_iB_{jk}.
\la{cs1}
\ee
\end{itemize}

In this paper we mainly use the scalar operator
$\phi^\dagger\phi$, the U(1) operator $B_{ij}$  
and the vector operator $W_i^3$.

\section{4d $\to$ 3d relation}
\label{4d3d}

As emphasized earlier, it is important to separate the two stages
of solving the finite $T$ problem: relating the 4d and 3d theories and
discussing the 3d theory as such. The mapping here is many $\to$ few;
many different 4d theories with different parameter values can 
correspond to a single set of 3d parameters $x,z$. This has to be
worked out for each case separately and here we only give a couple
of examples for illustration.

Firstly and most qualitatively, assume that we have an SU(2)$\times$U(1)
theory without fermions. Using tree-level relations between physics
and couplings and integrating out $A_0$, gives a simple 
analytic relation (eqs.(2.8-10) of \cite{nonpert}). According
to that relation, $x=0.06444,0.62402$ correspond to $m_H^*=60,180$ GeV and
$z=0,0.3$ correspond to $m_Z=m_W,m_Z^\rmi{exp}$.

Secondly, take the
minimal standard model and the 1-loop equations relating $x,y,z$
to physical parameters as given in \cite{generic}. 
Then $x=0.06444$ corresponds to $m_H=51.2$ GeV and 
$x=0.62402$ to $m_H=174$ GeV.
The value of $z=z_c$ at $T_c$ for given $m_H,
m_\rmi{top}$ is given in \fig\ref{zc}.
Since $z_c$ depends only logarithmically on $T$,
we have estimated $T_c$ from $y=0$ (the true $y_c$ is very close to zero,
see below). 

\begin{figure}[tb]
\vspace*{-2cm}
\hspace{1cm}
\epsfysize=15cm
\centerline{\epsffile{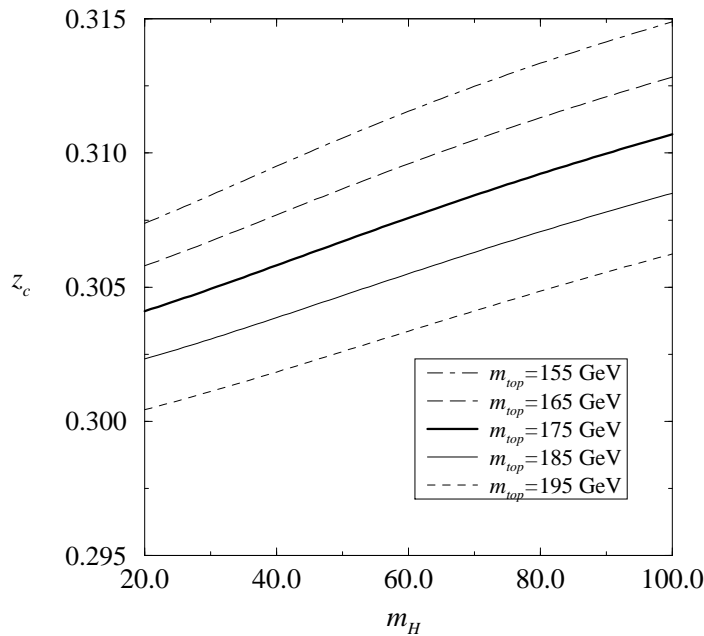}}

\vspace*{-5cm}
\caption[a]{\protect
The value of $z=g_3'^2/g_3^2=z_c$ at $T=T_c$
as a function of the physical Higgs
mass $m_H$ and the top quark mass $m_{\rm top}$ for the minimal
standard model. 
The dependence shown arises mainly from the dependence 
of $m_W$ on $m_H$, $m_{\rm top}$ (see~\cite{generic} for
the formulas used), and gives an 
estimate of the uncertainty in the physical value of $z_c$.
The value of $T_c$ is computed from $m_3^2(g_3^2)=0$; 
the dependence of $z_c$ on $T_c$ is only logarithmic.}
\la{zc}
\end{figure}

\begin{figure}[tb]
\vspace*{-2cm}
\hspace{1cm}
\epsfysize=15cm
\centerline{\epsffile{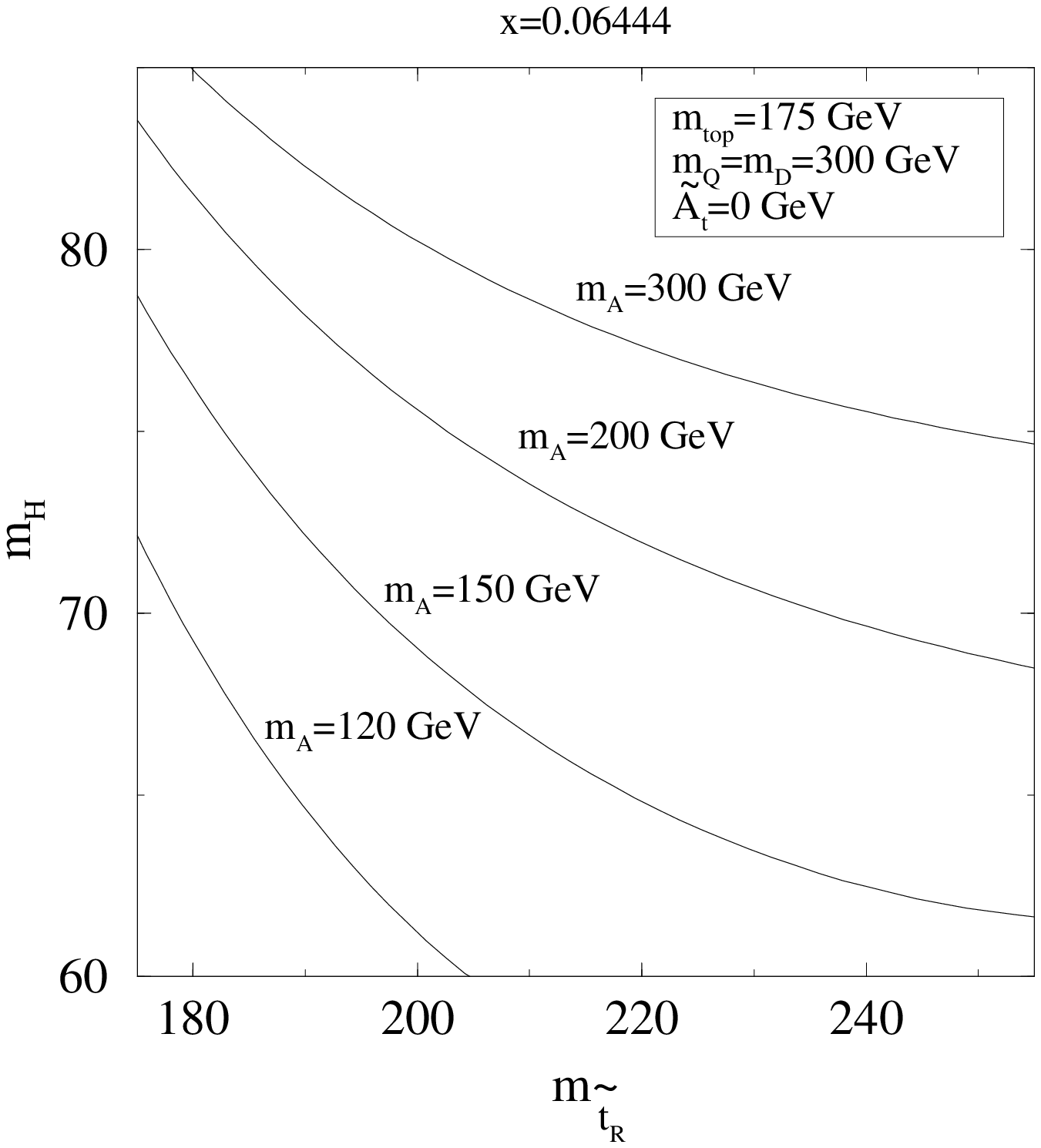}}
\vspace*{-6cm}
\caption[a]{\protect
Examples of parameter values corresponding to 
$x=0.06444$ in the MSSM. Here $m_{\tilde t_R}$
is the right-handed stop mass, $m_H$ is 
the lightest CP-even Higgs mass and $m_A$ is 
the CP-odd Higgs mass. The squark mixing parameters
have been put to zero.}
\la{mssm}
\end{figure}

As a third example, take the minimal supersymmetric standard model.
Fig.~\ref{mssm} shows different sets of values of the lightest Higgs
mass $m_H$, the right-handed stop mass $m_{\tilde t_R}$ and the 
CP-odd Higgs mass $m_A$ which all lead to the same value
$x=0.06444$.

\section{The lattice-continuum relations}

The lattice action corresponding to the continuum theory
(\ref{contaction}) is
\ba
S&=& \beta_G \sum_x \sum_{i<j}[1-\fr12 \tr P_{ij}] \nonumber \\
 &+& \beta_G' \sum_x \sum_{i<j}
[1-\fr12 (p_{ij}^{1/\gamma}+p_{ij}^{*1/\gamma})] \nonumber \\
 &-& \beta_H \sum_x \sum_{i}
\fr12\tr\Phi^\dagger(x)U_i(x)\Phi(x+i)
e^{-i\alpha_i(x)\sigma_3} \nonumber \\
 &+& \sum_x
(1-2\beta_R)\fr12\tr\Phi^\dagger(x)\Phi(x) + \beta_R\sum_x
 \bigl[ \fr12\tr\Phi^\dagger(x)\Phi(x)\bigr]^2.
\la{lagrangian}
\ea
Here the SU(2) and U(1) plaquettes are
\ba
P_{ij}(x) &=& U_i(x)U_j(x+i)
        U^\dagger_i(x+j)U^\dagger_j(x), \\
p_{ij}(x) &=&\exp\{i[\alpha_i(x)+\alpha_j(x+i)-
\alpha_i(x+j)-\alpha_j(x)]\}.
\ea
Any positive number $\gamma$ gives the same naive continuum
limit, but if $\exp(-i\alpha_i(x)\sigma_3)$,
$p_{ij}^{1/\gamma}$ are to be representations of the U(1) group
and $\alpha\in(0,2\pi )$, then
one has to choose $\gamma=1,1/2,1/3,\ldots$.
The topological effects associated with $\gamma\neq 1$ 
vanish in the continuum limit.

A crucially important relation is the one expressing the four
dimensionless lattice parameters $\beta_G$, $\beta_G'$,
$\beta_H$, $\beta_R$ of eq.~\nr{lagrangian} in terms of
the four dimensionless continuum parameters
$g_3^2a,x,y,z$. This relation, which can be derived using the
techniques in \cite{mlaine2}, is
\ba
g_3^2a & = & {4\over \beta_G}, \la{betag}\\
x & = & {\beta_R\beta_G\over\beta_H^2}, \la{betar}\\
z & = & \gamma^2\frac{\beta_G}{\beta_G'}, \la{z}\\
y & = &
\frac{\beta_G^2}{8\beta_H}
(1-2\beta_R-3\beta_H)
+{\Sigma\beta_G\over32\pi}
(3+12x+z)
\nonumber\\
&&+{1\over16\pi^2}\biggl[\biggl
({51\over16}-\frac{9}{8}z-\frac{5}{16}z^2
+9x-12x^2+3xz\biggr)
\biggl(\ln{\fr32\beta_G}+0.09\biggr) \nonumber \\
&&+
5.0-0.9z+\Bigl(0.01+\frac{1.7}{\gamma^2}\Bigr)z^2+5.2x+1.7xz\biggr],
\la{y}
\ea
where $\Sigma=3.17591$. This relation is exact in the limit
$a\to0$. In~\cite{moore_a} it was further
proposed (for $\gamma=1$)
that the 1-loop $O(a)$-corrections could be removed 
by modifying~\nr{betag}--\nr{z} with terms of relative
magnitude $O(1/\beta_G)$.

In usual discretisations of the U(1) gauge theory, $\gamma=1$.
In the physical situation the parameter $z$ (eq.(\ref{3dvariables}))
is about 0.3 and then the lattice coupling
$\beta_G'=\beta_G/z$ (eq.(\ref{z})) would be rather large when
approaching
the continuum limit $\beta_G\to\infty$. Using a value $\gamma<1$
permits one to have a smaller value of $\beta_G'$ which makes
finite size effects smaller; the price one pays is
an increasing correction term in eq.(\ref{y}) and larger
$O(a)$-corrections related to $g_3'^2$.

The historical motivation for the form (\ref{lagrangian}) of the
lattice action was to permit one to go to the fixed length limit of
the
Higgs field by taking $\beta_R\to\infty$. This limit is irrelevant
now and it is actually possible very simply to combine
(\ref{lagrangian})
and (\ref{betag})--(\ref{y}) by rescaling the Higgs field so that the
coefficient of the quartic term is +1. The lattice action for fixed
$a$
and fixed continuum variables $g_3^2,x,y,z$ then is
\ba
S\!\! &=& \!\! \beta_G \sum_x \sum_{i<j}[1-\fr12 \tr P_{ij}]
\nonumber \\
 &+& \!\! \beta_G' \sum_x \sum_{i<j}
[1-\fr12 (p_{ij}^{1/\gamma}+p_{ij}^{*1/\gamma})] \nonumber \\
 &-& \!\! \sqrt{{\beta_G\over x}} \sum_x \sum_{i}
\fr12\tr\Phi^\dagger(x)U_i(x)\Phi(x+i)
e^{-i\alpha_i(x)\sigma_3} \nonumber \\
 &+& \!\! \sqrt{{\beta_G\over x}} \biggl\{3+{8\over\beta_G^2}y
-{\Sigma(3+12x+z)\over4\pi\beta_G}-{1\over2\pi^2\beta_G^2}
\biggl[\ldots\biggr]
\biggr\} \sum_x \fr12\tr\Phi^\dagger(x)\Phi(x)
\nonumber \\
 &+& \!\! \sum_x
 \bigl[ \fr12\tr\Phi^\dagger(x)\Phi(x)\bigr]^2,
\la{lagrangiannew}
\ea
where the square bracket is the same as in (\ref{y}).

Finally, in the continuum limit,
the relation of the gauge-invariant lattice observable
$\langle\fr12\tr\Phi^\dagger\Phi\rangle$ in~\nr{lagrangian}
to the renormalized gauge-invariant continuum
quantity $\pp{\mu}$ in the $\msbar$ scheme is
\be
{\pp{\mu}\over g_3^2}=\frac{\beta_G\beta_H}{8}
\biggl\langle \fr12\tr\Phi^\dagger\Phi\biggr\rangle-
\frac{\Sigma\beta_G}{8\pi}-
{1\over16\pi^2}(3+z)\biggl(\log{\frac{3\beta_Gg_3^2}{2\mu}}
+0.67\biggr).\la{rl2}
\ee
The 1-loop $O(a)$ corrections to the discontinuity of $\pp{\mu}$
were also computed in~\cite{moore_a}.

\section{Perturbative results}

For understanding the magnitude of the expected effect of the
U(1) subgroup, it is useful to compute it in perturbation theory.
We have done this for the quantities listed in Section 2:
the critical curve $y=y_c(x,z)$,
the jump $\Delta\ell_3$ of the order parameter
$\ell_3\equiv\langle\phi^\dagger\phi(g_3^2)\rangle/g_3^2$ between
the broken (b) and symmetric (s) phases at $y_c$ and
the interface tension $\sigma_3$.

The coarsest approximation, which nevertheless gives the general
pattern and can be given in analytic form, is obtained by taking the
1-loop potential and including only the vector loops. The result
then is
\ba
y_c(x,z)&=&{1\over128\pi^2x}\biggl[\fr23+\fr13(1+z)^{3/2}\biggr]^2,
\nonumber\\
\Delta\ell_3&=&{1\over128\pi^2x^2}
\biggl[\fr23+\fr13(1+z)^{3/2}\biggr]^2,
\nonumber\\
\sigma_3&=&{1\over6(16\pi)^3}\biggl({2\over x}\biggr)^{5/2}
\biggl[\fr23+\fr13(1+z)^{3/2}\biggr]^3.
\ea
Here one sees concretely how increasing $z$ increases the quantities
studied.

A more accurate result is obtained with 2-loop optimised
3d perturbation theory. The effective potential for the 3d SU(2)+Higgs
theory is given and its optimisation is discussed in
\cite{perturbative}. Since the effects of $g_3'^2$ are very small,
it is sufficient for the present purpose
to differentiate between $m_W^2$ and $m_Z^2$
only at 1-loop level (the complete 2-loop potential can
be inferred from the 4d results in \cite{ae,fh}). A numerical
computation then leads to the results shown in \figs\ref{xyz}, \ref{Dl3}.
The figures also give the results from lattice Monte Carlo
simulations \cite{nonpert,isthere?} and make it clear how
the perturbative discussion is a good qualitative quide, but fails
in a crucial property of the transition: its termination.

\begin{figure}[tb]

\vspace{-1cm}

\hspace{1cm}
\epsfysize=15cm
\centerline{\epsffile{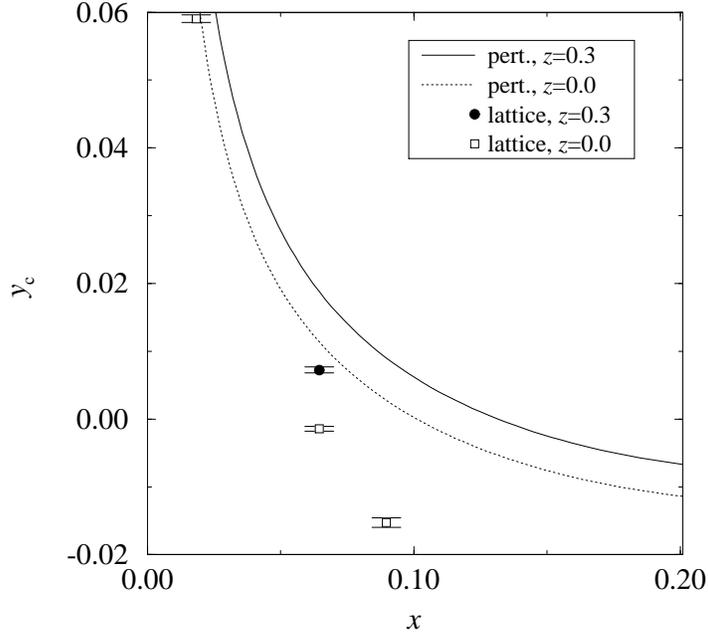}}

\vspace{-5cm}

\caption[a]{\protect The curves show the values of $y_c(x,z)$
computed from the 3d 2-loop effective potential of 
the SU(2)$\times$U(1)+Higgs model.
Lattice Monte Carlo results are also shown. The perturbative curves 
continue to large values of $x$, but on the lattice the first
order transition ends at $x\sim 1/8$. Beyond that in the cross-over
region, the values obtained for $z=0$
with the max($\chi_{R^2}$)-method (see Sec.~6.1)
are $y_c(0.274,0)=-0.066,\,y_c(0.624,0)=-0.16$.}
\la{xyz}
\end{figure}
\begin{figure}[tb]

\vspace*{-1cm}
\centerline{\hspace{-3.3mm}
\epsfxsize=10cm\epsfbox{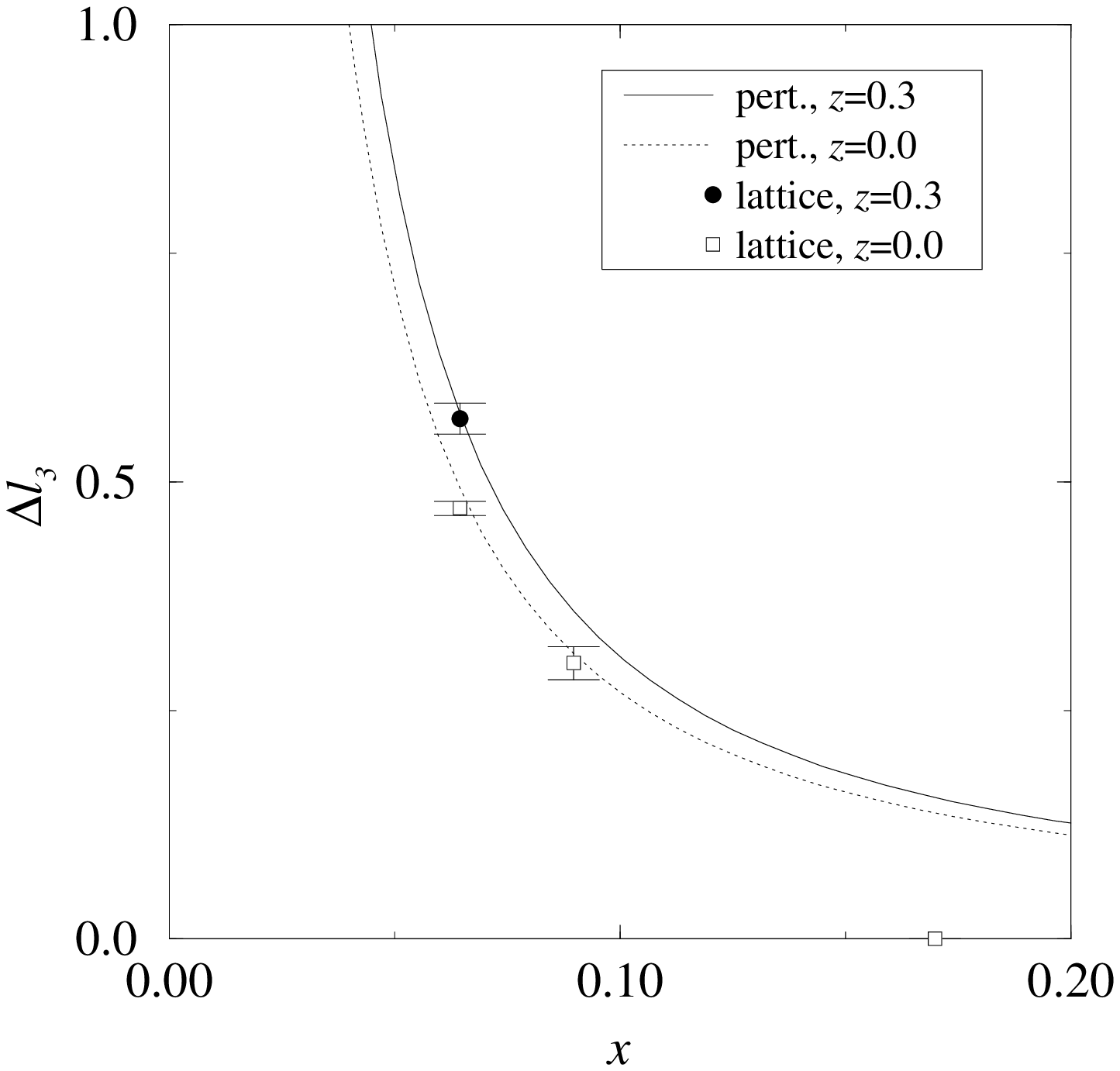}
\hspace{-2cm}
\epsfxsize=10cm\epsfbox{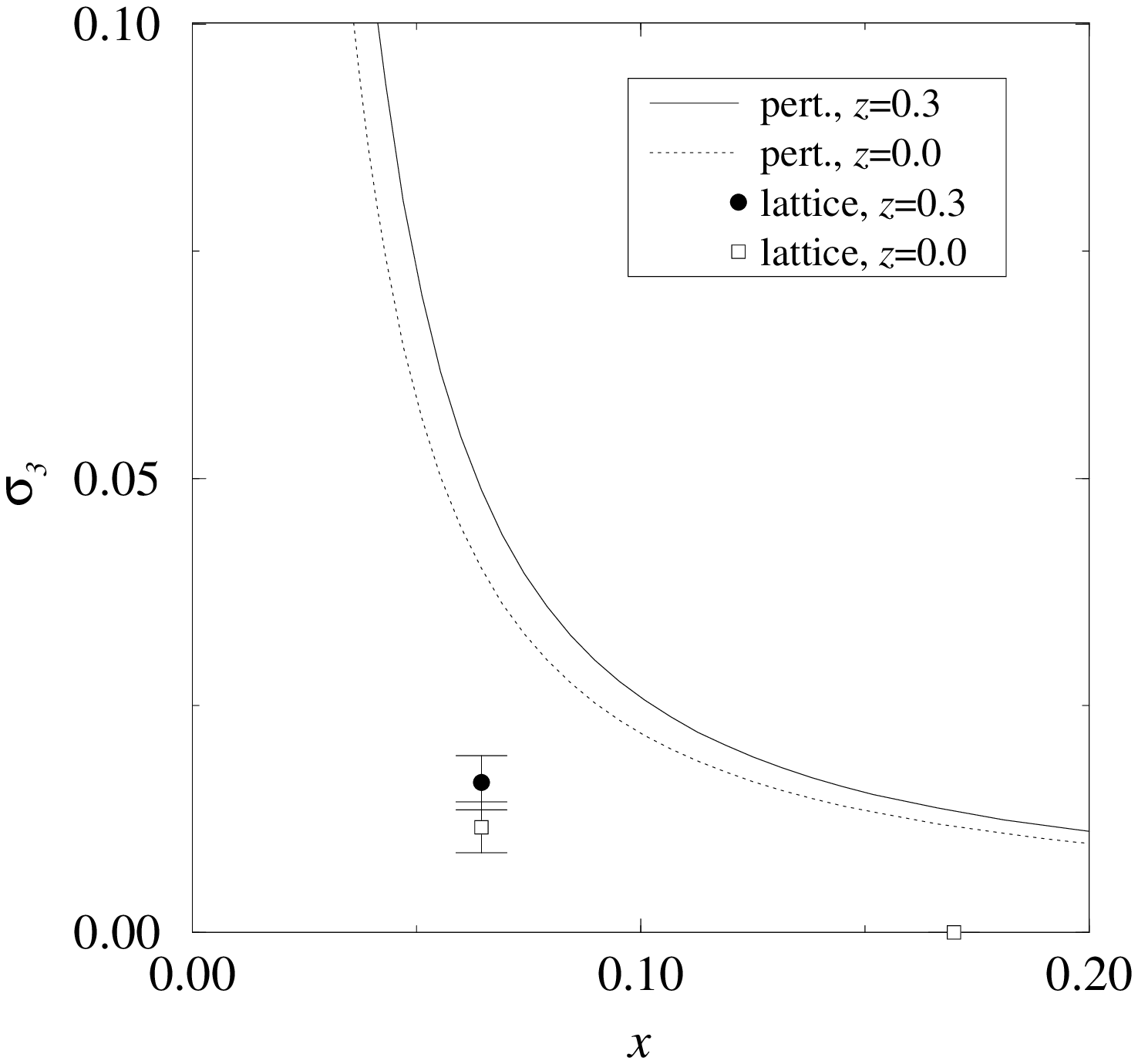}}

\vspace*{-5cm}
\caption[a]{The same as \fig\ref{xyz} but for the jump of the order
parameter and the interface tension. At $x\approx 0.096$, the interface
tension was estimated in~\cite{leip} to be 
$\sigma_3\sim 1\times 10^{-3}$. For $x\gsim 1/8$, there is 
no transition and $\Delta\ell_3$, $\sigma_3$ vanish.}
\la{Dl3}
\end{figure}

\section{Simulations}

The simulations with the SU(2)$\times$U(1)+Higgs action
(\ref{lagrangian}) are rather similar to those for the SU(2)+Higgs
theory, described in detail in \cite{nonpert}.  We shall, therefore,
mainly just present the results and compare them.  We have performed
simulations with $x = 0.06444$ and $x=0.62402$.  For the SU(2)+Higgs
case the former value gives a clear first order transition, whereas
for the latter value the transition becomes a regular cross-over
\cite{nonpert,isthere?}.
Since the most precise results for the SU(2)+Higgs theory were for
$x=0.06444$ we shall concentrate mostly on this $x$-value.  For $z$
we always take $z=0.3$.

The simulations were mainly performed on a Cray C-90 supercomputer at
the Center for Scientific Computing, Finland.  Some of the smaller
volume simulations were done on workstation clusters.  The total
amount of computing power used for the $z=0.3$ case was about 4.5 Cray
cpu-months, or $2.3\times 10^{15}$ floating point operations, to be
compared with $5\times 10^{15}$ flop for $z=0$.

\subsection{The transition point $y=y_c(x=0.06444,z=0.3)$}

\begin{figure}[tb]
\figtopspace
\centerline{\epsfysize=\figysize \epsfbox{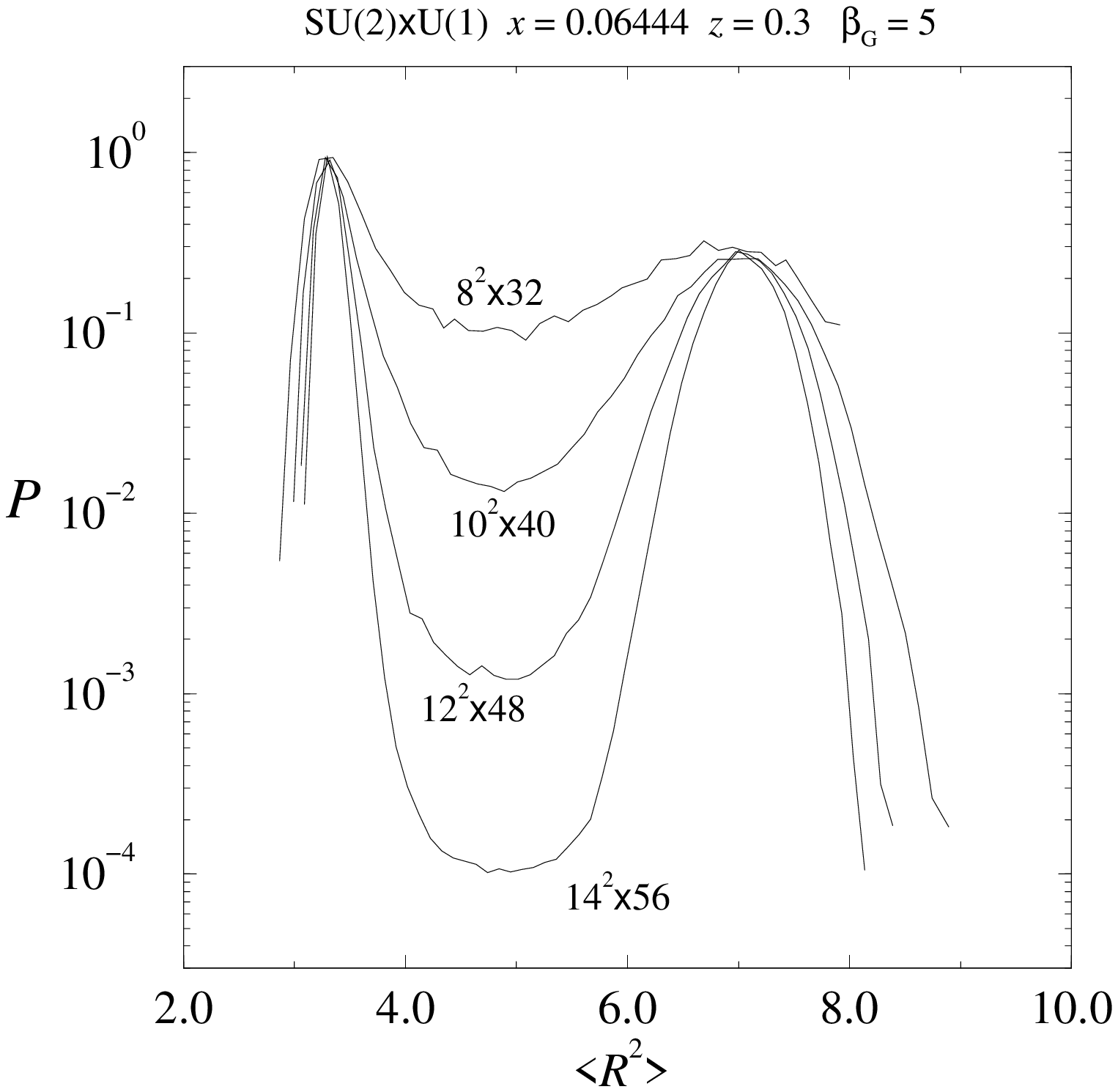}}
\vspace*{-6cm}
\centerline{\epsfysize=\figysize \epsfbox{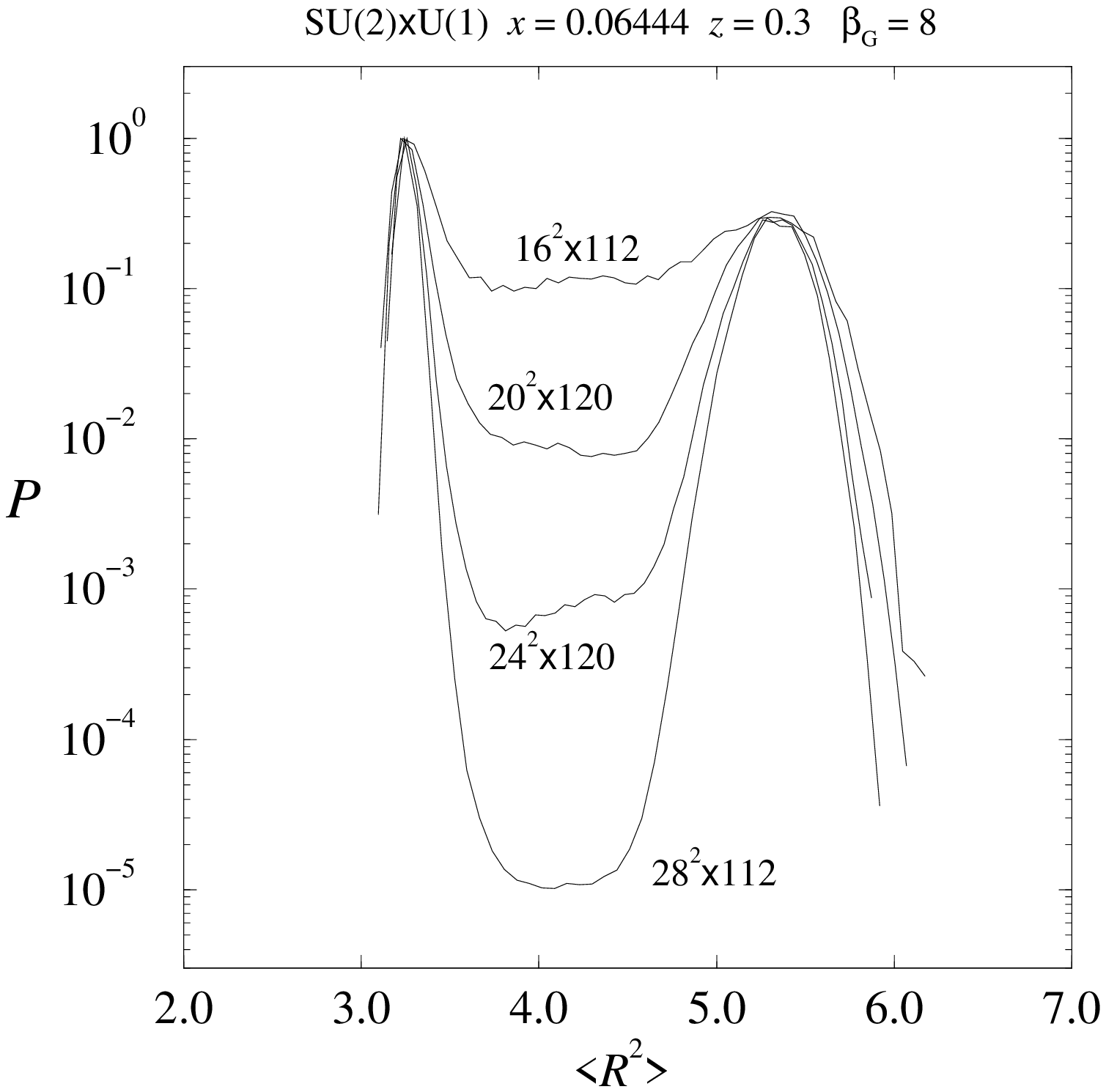}}
\figbottomspace
\caption[a]{The ``equal weight'' distributions of the average Higgs
length
squared $R^2 = \sum_x R^2(x)/V$
at the transition point for $\beta_G = 5$ (upper) and
8 (lower) for $x=0.06444$.}
\label{m60_R2histograms}
\end{figure}

Let us first discuss the $x=0.06444$ case, where the transition is
relatively strongly first order ($m_H^* = 60$\,GeV in the notation
used in \cite{nonpert}).  The SU(2)$\times$U(1)+Higgs model does not
have a local gauge invariant order parameter.  We identify the
transition by studying order parameter like quantities, $R^2 \equiv
\sum_x R^2(x)/{\rm Vol} $ and
\be
  L \equiv \frac{1}{3{\rm Vol}} \sum_{x,i}
        \fr12 \tr V^\dagger(x) U_i(x) V(x+i)
        e^{-i\alpha_i(x)\sigma_3}\,,
\label{Loperator}
\ee
where $V$ is the SU(2)-direction of the Higgs field: $\Phi = RV$.
These operators develop a discontinuity at the transition point, as
shown by the probability distributions $p(R^2)$ in
\fig\ref{m60_R2histograms} for $\beta_G =5$ and 8.  The development
of a strong two-peak structure with an increasing volume is
unambiguous.

In order to obtain the physical value of $y_c(x,z)$ we have to treat
the finite volume effects and finite lattice spacing effects
systematically.  Our procedure here consists of the following stages:

{\bf The pseudocritical coupling.~}
For fixed $x$ and $\beta_G$ (lattice spacing) and each lattice size
we determine the pseudotransition coupling $\beta_{H,c}$ by several
methods:

 (1) maximum location of the $R^2$-susceptibility
        $\chi_{R^2} = \langle(R^2 - \langle R^2\rangle)^2\rangle$,

 (2) minimum location of the Binder cumulant
        $B_L = 1 - \langle L^4\rangle /(3\langle L^2\rangle^2)$,

 (3) ``equal weight'' $\beta_H$-value for the distribution $p(R^2)$,

 (4) ``equal height'' $\beta_H$-value for $p(L)$.

\noindent
In practice, we perform a series of simulations with different values
of $\beta_H$ for fixed $\beta_G$ and $x$ (implying {\em not} fixed
$\beta_R$ but given by eq.(\ref{betar})) until we have a good 
coverage of the transition region.
The $\beta_H$-values corresponding to the above criteria are then
found with the Ferrenberg-Swendsen multihistogram reweighting
\cite{Ferrenberg88}, and the error analysis is performed
with the jackknife method.  For all but the smallest volumes it
is necessary to use multicanonical simulations
\cite{multicanonical}.
For technical details we refer to \cite{nonpert}, where a similar
analysis was carried out for the SU(2)+Higgs model.

\begin{table}[ht]
\newcommand{\tube}[2]{~~$#1^2\times #2$}
\newcommand{\cube}[1]{~~$#1^3$}
\newcommand{\mc}{_m}
\center
\begin{tabular}{|c|c|c|lll|} \hline
 $x$ & $\beta_G$ & $\gamma$ & \multicolumn{3}{c|}{volumes}  \\
\hline\hline
 0.06444 &  5 & 1 & \tube{8}{32\mc}&  \tube{10}{40\mc} & \\
         &    &   & \tube{12}{42\mc}& \tube{14}{56\mc} &
        \\ \cline{2-6}
         &  8 & 1 &  \multicolumn{1}{l}{
        \cube{24\mc}\,\, \cube{32\mc}} &
\tube{16}{112\mc}&\tube{20}{120\mc}\\
         &    &   &  \tube{24}{120\mc}& \tube{28}{112\mc} &
        \\ \cline{2-6}
         & 16 &1/2&  \multicolumn{3}{l|}{
        \cube{12}\,\,\cube{16}\,\,\cube{24}\,\, \cube{32}} \\
         &    &   & \tube{24}{72\mc}& \tube{32}{96\mc}&
\tube{40}{120\mc}
        \\  \hline
 0.62402 &  8 & 1 & \multicolumn{3}{l|}{
        \cube{12}\,\,\cube{16}\,\,\cube{24}\,\,\cube{32}}
        \\  \hline\hline
\end{tabular}
\caption[0]{The lattice sizes used for the simulations at the transition
temperature for each ($x,\beta_G$)-pair.   In most of the cases,
several $\beta_H$-values were used around the transition point.
Multicanonical simulations are marked with the subscript
($\mc$).}\la{table:lattices}
\end{table}

The lattice volumes used in the simulations are listed in Table
\ref{table:lattices}.  For $x=0.06444$ we used $\beta_G$ values 5, 8
and 16\@; in the last case we used $\gamma=1/2$ in order to avoid
too large finite size effects.

\begin{figure}[tb]
\figtopspace
\epsfysize=\figysize
\centerline{\epsffile{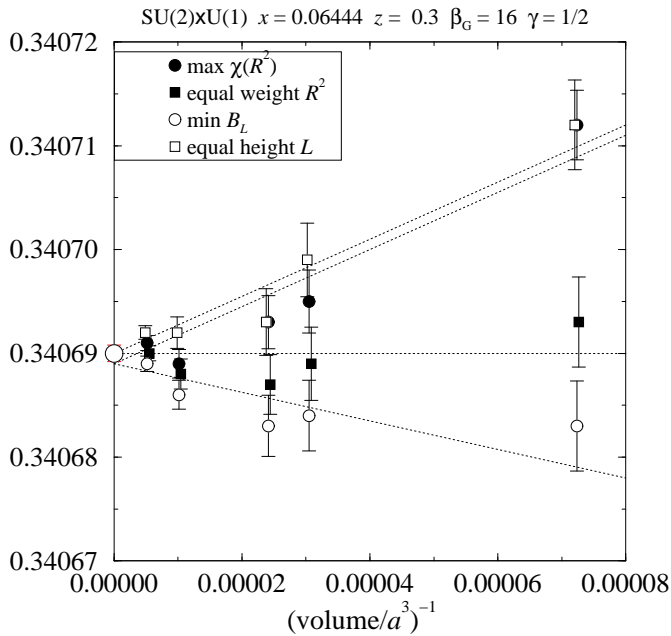}}
\figbottomspace
\caption[a]{The $V=\infty$ limit of the pseudocritical couplings
$\beta_{H,c}$ for $x=0.06444$ ($m_H^*=60$~GeV) and $\beta_G=16$,
calculated with four different methods. The figures for 
$\beta_G=5, 8$ are similar.}
\la{um60b16-bc}
\end{figure}


\begin{table}[ht]
\center
\begin{tabular}{|r|c|l|l|}
\hline
  $\beta_G$ & $\gamma$ &\cen{$\beta_{H,c}$} &\cen{$y_c$} \\
\hline
      5 & 1   & 0.359863(3)   & 0.00199(7)   \\
              8 & 1   & 0.3489001(11) & 0.00255(7)   \\
             16 & 1/2 & 0.3406899(8)  & 0.00489(20)  \\
          $\infty$ &  & 1/3           & 0.0072(5)    \\
\hline
\end{tabular}
\caption[a]{The infinite volume critical couplings
$\beta_{H,c}$ and the associated values of $y_c$ 
at $x=0.06444,\,z=0.3$ (from
eq.(\ref{y})). The values of $\beta_{H,c}$ are calculated from the
``equal weight of $p(R^2)$'' data.\la{table:betac}}
\end{table}

{\bf The infinite volume limit.~} For any given lattice, the criteria
(1)--(4) above yield different values for the pseudocritical coupling
$\beta_{H,c}$.  However, in the thermodynamic limit
$V\rightarrow\infty$, all the methods extrapolate very accurately to
the same value, as shown in \fig\ref{um60b16-bc}.  It should be noted
that the different methods do not yield statistically independent
results, and combining the results together is not justified.  In the
extra\-polations we used only volumes which are large enough in order to
be compatible with the linear behaviour in $1/V$.  In
Table~\ref{table:betac} we give the results using the ``equal weight''
values of the $p(R^2)$ distributions.  Also shown are the
corresponding values of $y_c$, obtained through
\eq\nr{y}.  Note that the value $\gamma=1/2$ for $\beta_G=16$ enters
only through a 2-loop contribution in \eq\nr{y}; nevertheless, its
inclusion is numerically crucial.

\begin{figure}[tb]
\figtopspace \vspace{1cm}
\epsfysize=\figysize
\centerline{\epsffile{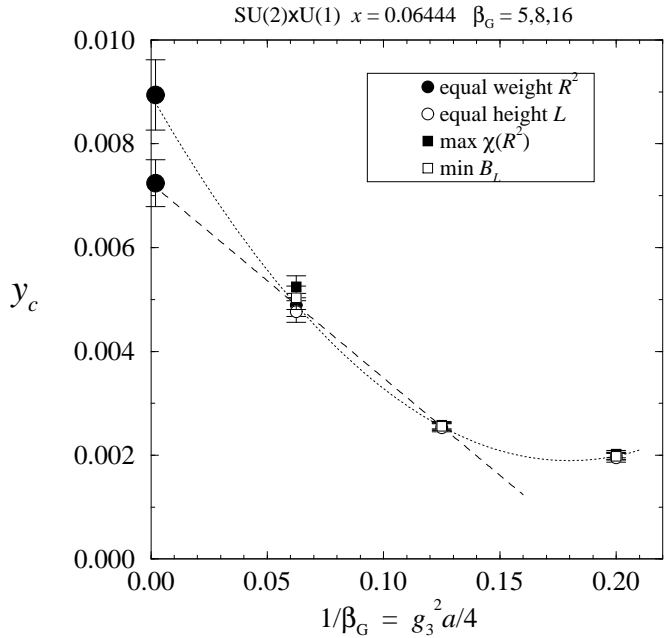}}
\figbottomspace
\caption[a]{The continuum limit ($\beta_G\rightarrow\infty$) of the
critical temperature for $x=0.06444$ ($m_H^*=60$~GeV), $z=0.3$.}
\la{um60_yc}
\end{figure}

{\bf The continuum limit.~}
Lastly, the physical value of $y_c$ is
obtained by extrapolating the $y_c(x,z;\beta_G)$ values to zero lattice
spacing ($\beta_G=4/(g_3^2a)\rightarrow\infty$).  This is shown in
\fig\ref{um60_yc}.  The leading behaviour is linear in $a$; however,
it is clearly impossible to fit a straight line through all
$\beta_G$-values.  In principle, the finite $a$ curve is {\em
different\,} for each value of $\gamma$, and fitting an extrapolation
through $\beta_G=5$ and 8 ($\gamma=1$) and $\beta_G=16$
($\gamma=1/2$)
datapoints is not rigorously justified.  Nevertheless, when we
compare
this curve with the SU(2)+Higgs one (\fig11 in \cite{nonpert}), the
qualitative similarity is obvious.  
However, the curvature of the quadratic fit is here
clearly larger, and at $\beta_G=5$ the quadratic fit already starts to
turn upwards.  We assume that this behaviour is not physical
and that the quadratic extrapolation overestimates the continuum
value
of $y_c$.  Therefore, in Table~\ref{table:betac} we quote the value
extrapolated linearly from $\beta_G=8$ and $\beta_G=16$ -points;
however, the difference between the linearly and quadratically
extrapolated values is small and does not affect our conclusions.

In Table~\ref{table:results} we compare the Monte Carlo results
with the perturbative values. It is seen that for $\Delta\ell_3$
and $\sigma_3$ (whose determination is discussed below)
the effect of $z$ is just what is expected from 
the dominant 1-loop perturbative effect of $g_3'^2$. 
For $y_c$ the discrepancy is slightly larger; however,
here the lattice determination of $y_c$ is not that 
easy (as explained in the previous paragraph). In any case
the difference is small, and
in terms of the critical temperature $T_c$ the discrepancy is
below 1\% since $y_c$ is composed of $m_H^2$ and $g^2 T_c^2$
as a difference of two large terms. Thus for any 
physical conclusions, we can say that the last block 
in Table~10 of~\cite{nonpert}, which was based on a purely
perturbative estimate of the effects of the U(1) group, 
can be considered reliable.

\begin{table}[ht]
\center
\begin{tabular}{|l|l|l|l|}
\hline
 measured   & $z=0$ & $z=0.3$  & expected if U(1) perturbative \\
\hline
$y_c^\rmi{latt}$             & -0.00142(36) & 0.00724(45) & 0.0060 \\
$y_c^\rmi{2-loop}$           &  0.01141     & 0.01882     &   \\
$\Delta\ell_3^\rmi{latt}$    &  0.471(8)    & 0.569(17)   & 0.55  \\
$\Delta\ell_3^\rmi{2-loop}$  &  0.493       & 0.575       &   \\
$\sigma_3^\rmi{latt}$        &  0.0116(28)  & 0.0165(30)  & 0.014  \\
$\sigma_3^\rmi{2-loop}$      &  0.0401      & 0.0487      &   \\
\hline
\end{tabular}
\caption[a]{Comparison of measured values and expected values at
$x=0.06444, z=0.3$ if the effect of U(1) is perturbative. 
The 'expected' values
are obtained by computing the difference ($y_c$) or ratio
($\Delta\ell_3$,
$\sigma_3$) of the lattice and perturbative values at $z=0$,
and by using these numbers in modifying the perturbative
values for $z=0.3$. Note that the $z=0$ values are slightly
different from those in Table~10 of~\cite{nonpert}, since
there the extrapolation to the continuum limit 
was made using the 4d observables
$T_c^*$, $L/T_c^{*4}$, $\sigma/T_c^{*3}$ whereas here
we use directly the 3d observables.\la{table:results}}
\end{table}

\begin{figure}[tb]
\figtopspace
\centerline{\epsfysize=\figysize \epsfbox{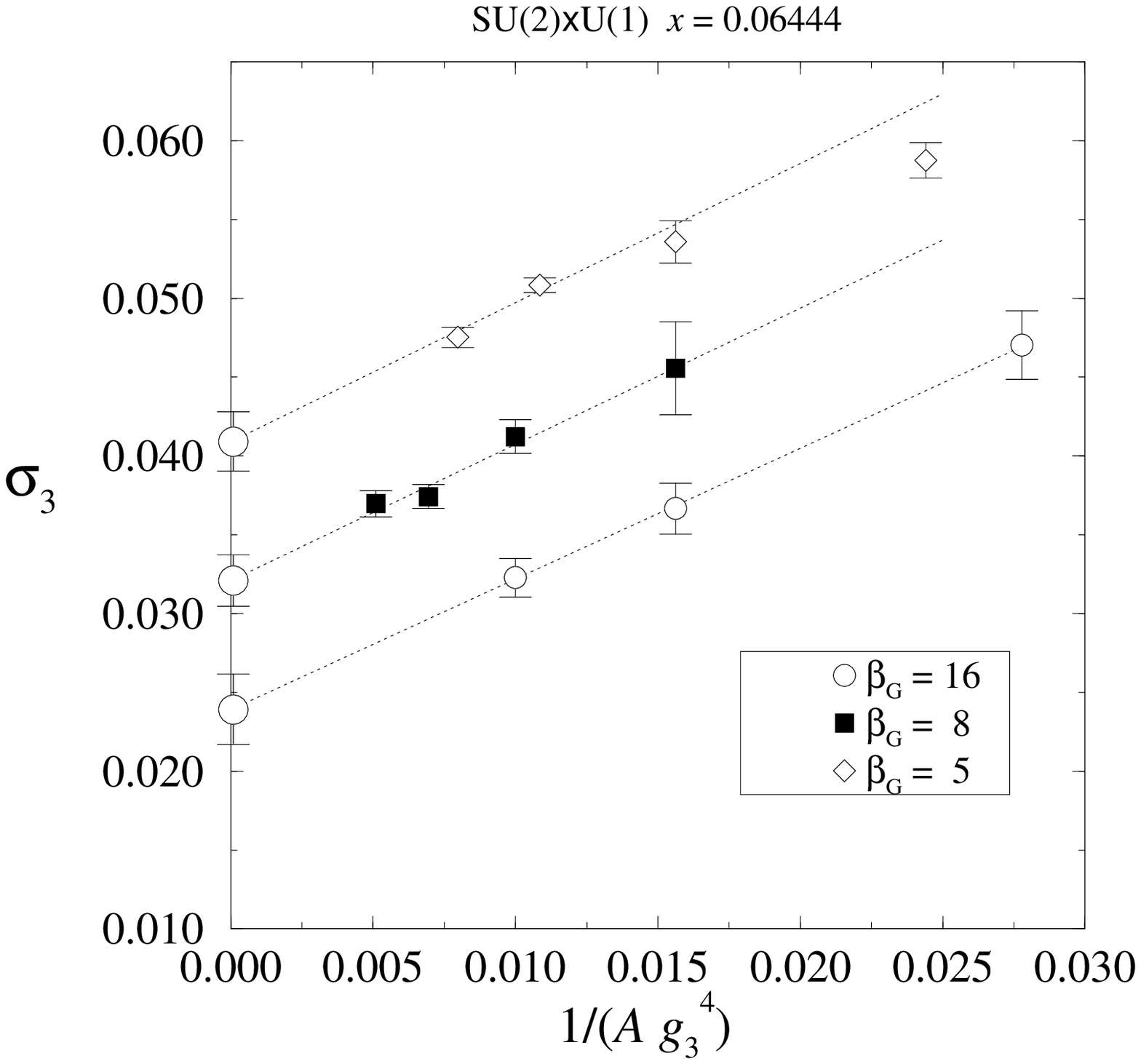}}
\vspace*{-6cm}
\centerline{\epsfysize=\figysize \epsfbox{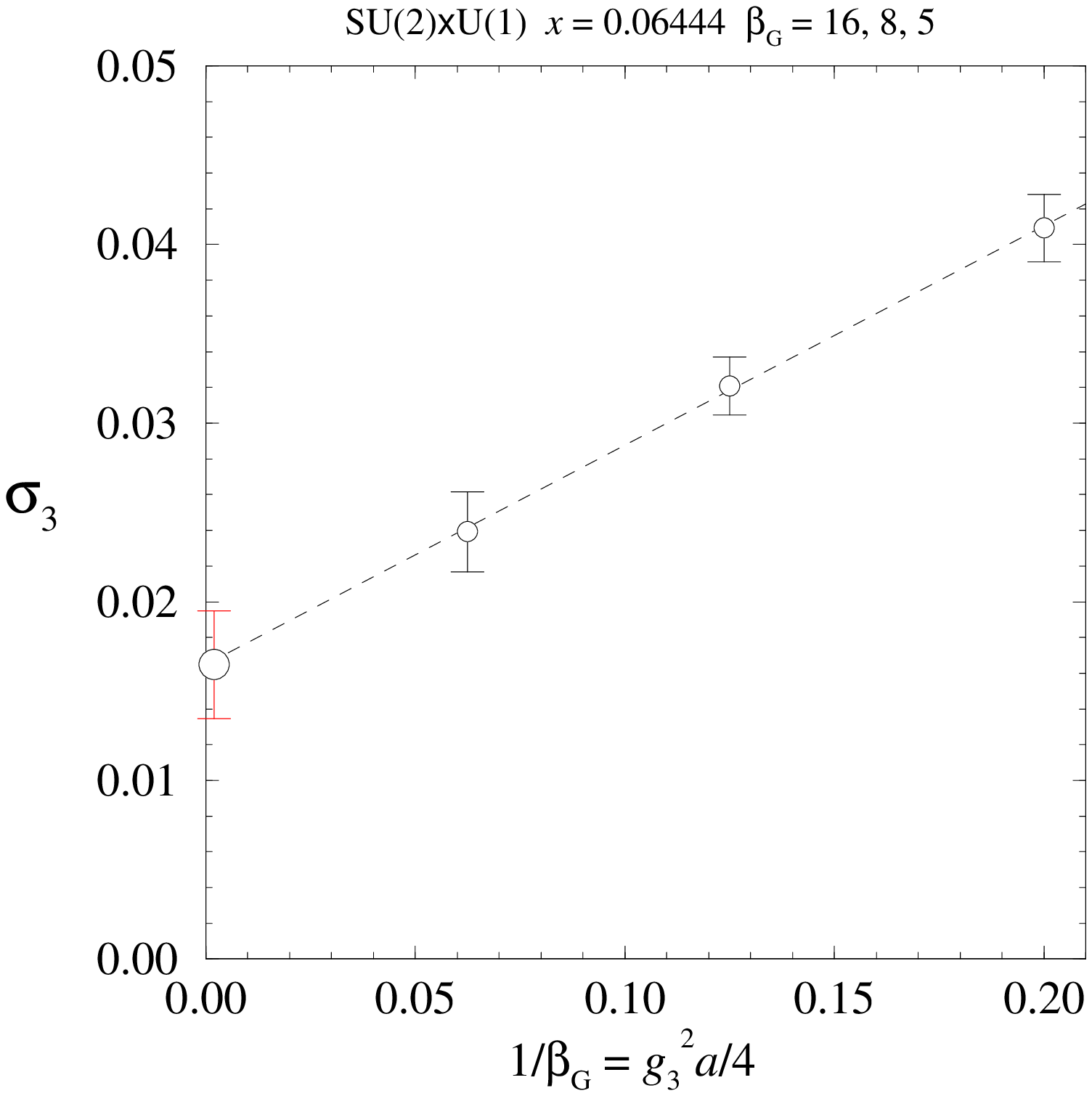}}
\figbottomspace
\caption[a]{The interface tension $\sigma_3$ extrapolated to infinite
volume (top) and then to the continuum limit (bottom) for $x=0.06444$,
$z=0.3$.}
\label{um60sigma}
\end{figure}

\begin{figure}[tb]
\figtopspace \vspace{1cm}
\epsfysize=\figysize
\centerline{\epsffile{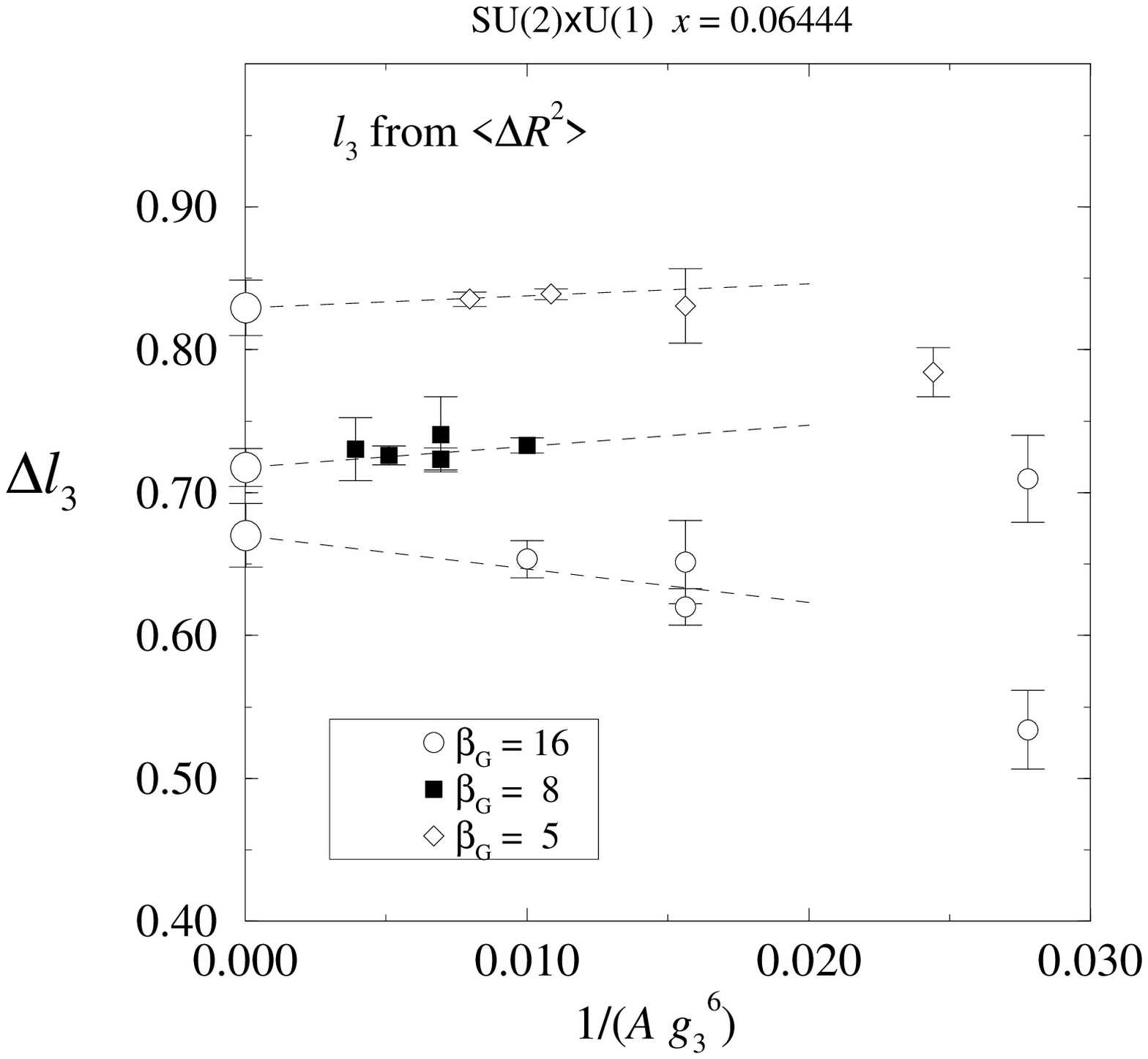}}
\figbottomspace
\caption[a]{The infinite volume limit of the discontinuity
$\Delta\pdp/g_3^2$
for $x=0.06444$, $z=0.3$.  The extrapolation is done linearly in
$1/A$, where $A$ is the smallest cross-sectional area of the
system.\la{um60latent}}
\end{figure}

\subsection{The interface tension and $\Delta\pdp$}

We measure the interface tension with the histogram method
\cite{Binder82}.
At the transition point the probability of the two bulk phases is
equal, but a system of a finite volume can also exist in a mixed
state
consisting of domains of the two bulk phases.  The probability of the
mixed state is suppressed by the free energy of the interface between
the bulk phases.  As a result, an order parameter distribution
develops a characteristic 2-peak structure
(\fig\ref{m60_R2histograms}),
and the interface tension $\sigma_3$ can be extracted from the
limit
\be
\sigma_3 = \lim_{V\to \infty}\frac{1}{2g_3^4A}\ln
\frac{P_{\rm max}}{P_{\rm min}},
\ee
where $P_{\rm max}$ and $P_{\rm min}$ are the values of the
distribution
at the peak and at the plateau between the peaks, and $A$ is the
smallest cross-sectional area on a three-dimensional periodic volume.
In practice a careful finite size analysis is required; for technical
details we again refer to the discussion in \cite{nonpert}.

The extrapolation of $\sigma_3$ to the infinite volume for each
lattice spacing is shown in the upper part of
\fig\ref{um60sigma} and to the continuum limit $a=0$ in the bottom
part.
The final result is shown in Table~\ref{table:results} together
with the perturbative estimates.

Let us next consider the measurements of the discontinuity of the
condensate $\pdp$.  The lattice quantities
$R^2\equiv\half\tr\Phi^\dagger\Phi$ are first converted to the values
of the continuum condensate at scale $g_3^2$ by using \eq\nr{rl2}:
$\ell_3\equiv\langle\phi^\dagger\phi(g_3^2)\rangle/g_3^2$.  The
$\Delta \langle R^2\rangle$ discontinuity is extracted from the
``equal weight'' distributions $p(R^2)$ by measuring the positions of
the peaks.  The peak positions are determined by fitting parabolas to
the distributions near the peaks.  For each value of $\beta_G$ the
measurements are extrapolated to infinite volume with respect to
$1/A$
(\fig\ref{um60latent}), and these values are in turn extrapolated to
the continuum limit.  The final results are shown in
Table~\ref{table:results}.

An alternative method for calculating $\Delta\pdp$ is to use 
\eq\nr{contaction} and measure
\be
  \fr{\Delta\pdp}{g_3^2} = \fr{-1}{Vg_3^6} \fr{d\Delta\log Z}{dy}
                         = \fr{-1}{Vg_3^6} \fr{d\Delta P}{dy},
\ee
where $\Delta P$ is the difference of the probabilities of the
two phases and the derivatives are evaluated at the transition
point (where $P_{\rm symm.} = P_{\rm broken} = 1/2$).  The result
obtained with this method is quite compatible with the above one.

\subsection{Higgs and $W$ correlators}

\begin{figure}[tb]
\figtopspace \vspace{1cm}
\epsfysize=\figysize
\centerline{\epsffile{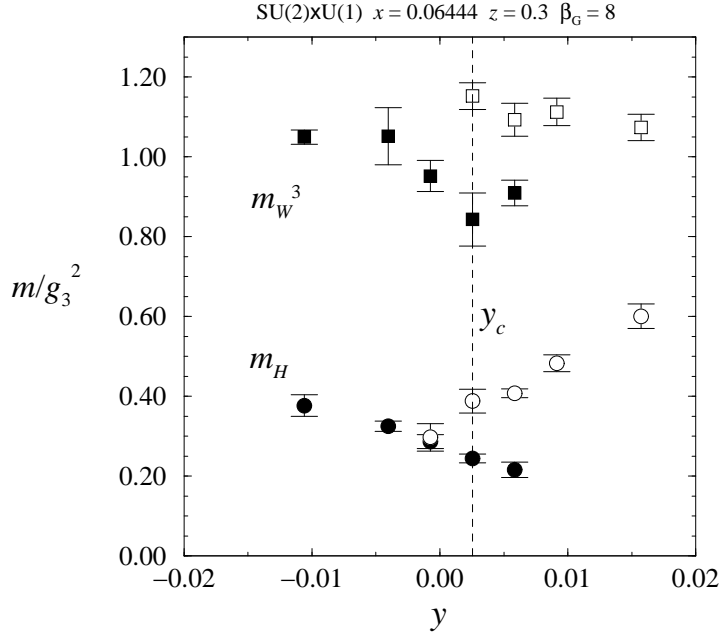}}
\figbottomspace
\caption[a]{$m_H$ and $m_{W^3}$ measured from a $\beta_G=8$,
volume $=32^2\times 64$ -lattice.  Filled symbols correspond to broken phase,
open symbols to symmetric phase masses.\la{um60mhmw}}
\end{figure}

We measure the masses of $H$ and $W^3$ with the lattice versions of
the operators $\phi^\dagger\phi$ and $\tr\Phi^\dagger
iD_i\Phi\sigma_3$, respectively.  The correlation functions are
measured in the direction of the $x_3$ -axis, and in order to enhance
the projection to the ground states, we use blocking in the
$(x_1,x_2)$-plane.  The fields are recursively mapped from blocking
level $(k) \rightarrow (k+1)$ ($\Phi^{(k)}(x) \rightarrow
\Phi^{(k+1)}(y)$), so that the fields on the $(k+1)$-level lattice are
in effect defined only on the even points of the $(k)$-level lattice on the
$(x_1,x_2)$-plane, doubling the lattice spacing along the
$(x_1,x_2)$-plane: $(a_1,a_2,a_3)^{(k+1)} = (2a_1,2a_2,a_3)^{(k)}$.
The blocking is performed with the transformations
\be
  \Phi^{(k+1)}(y) = \fr15 \Phi^{(k)}(x) +
        \fr15 \sum_{i=\pm 1,2} U^{(k)}_i(x) \Phi^{(k)}(x+i)
                                \exp[-i\alpha^{(k)}_i(x)\sigma_3]
\ee
and ($i = 1,2, j\neq i$)
\ba
  U^{(k+1)}_i(y) &=& U'^{(k)}_i(x)U'^{(k)}_i(x+i), \la{ublock1} \\
  U'^{(k)}_i(x)  &=&  \fr13 U^{(k)}_i(x)  +  \fr13 \sum_{j=\pm 1,2}
        U^{(k)}_j(x) U^{(k)}_i(x+j)
                U^{(k)\dagger}_j(x+i), \la{ublock}
\ea
where $(x_1,x_2,x_3) \equiv (2y_1,2y_2,y_3)$ and $U_{-i}(x) =
U_i^\dagger(x-i)$.  The U(1) gauge field $B_i^{(k)} \equiv
\exp[-i\alpha^{(k)}_i\sigma_3]$ is blocked with a transformation
similar to (\ref{ublock1}--\ref{ublock}).  With the blocked fields we
can define the operators
\ba
  h^{(k)}(z)    &=& \sum_x \delta_{x_3,z}
                \tr\Phi^{(k)\dagger}(x)\Phi^{(k)}(x), \\
  w_i^{3(k)}(z) &=& \sum_x \delta_{x_3,z}
                \tr\Phi^{(k)\dagger}(x)
                        U_i^{(k)}(x)\Phi^{(k)}(x+i)
                        \exp[-i\alpha^{(k)}_i(x)\sigma_3]\sigma_3.
\ea
These operators are used to calculate the zero transverse 
momentum correlation functions
\ba
  H^{(k)}(l)  &=& \fr{1}{V} \sum_{z} h^{(k)}(z) h^{(k)}(z+l), \\
  W^{3(k)}(l) &=& \fr{1}{2V} \sum_{z;\,i=1,2}
                    w_i^{3(k)}(z) w_i^{3(k)}(z+l),
\ea
and the masses are measured from the exponential fall-off in these
functions.  The blocking method used here resembles the
one used by Philipsen \etal \cite{phtw}; however, we do not perform
the diagonalization of the full cross-correlation matrix formed
of operators of different blocking levels.

For the Higgs channel the blocking makes little difference, and for
the final results we use non-blocked $(k=0)$ operators.  For $W^3$, on
the other hand, the blocking improves the signal substantially, and we
use 3 times blocked operators $W^{3(3)}$ to measure the $W^3$ mass.
At $x=0.06444$ we measure the masses using $\beta_G=8$, $32^2\times
64$ lattices and performing simulations on both sides of the
transition.  The metastability of the transition is large enough to
permit the mass measurements also in the metastable branches, as shown
in
\fig\ref{um60mhmw}.  This figure can be compared with the SU(2)+Higgs
masses in Fig.~16 of \cite{nonpert}; it should be noted that 
there the $y$ range $-0.066<y<0.042$ around the critical coupling 
$y_c$ is much wider than in \fig\ref{um60mhmw}.

\subsection{The photon screening mass} 
\la{photon}

The qualitative difference between the
SU(2)$\times$U(1) theory considered in this paper and 
the SU(2)$+$Higgs model studied earlier is the presence of an extra gauge
boson, associated with the U(1) group. This particle is massless in
all orders of perturbation theory. 

There are several questions which can be addressed here. The first
one is associated with the very existence of the massless vector
excitation. Indeed, there are known examples when the photon gets
a mass due to some non-perturbative effects. For example, in 
compact (discretized)
QED in three dimensions the photon in fact does not exist
and is replaced by a massive pseudo-scalar 
excitation~\cite{polyakov}. 
If some non-perturbative photon mass is
generated, then, as has been argued in \cite{lindemon}, the
problem of the primordial magnetic monopoles would be solved.

The second question is associated with $Z-\gamma$ mixing. At
sufficiently large and positive $y$ (high temperatures) one would say
that the massless excitation corresponds mainly to the hypercharge
field $B$, while at large negative values of $y$ (small temperatures)
it is a mixture of the hypercharge field and the third component of the
SU(2) gauge field. Since the SU(2)$\times$U(1) theory does not
contain a local gauge-invariant 
order parameter distinguishing the ``broken'' and
``symmetric'' phases, at sufficiently large values of the parameter
$x$ (Higgs mass), the ``content'' of the hypercharge field in the
massless state should interpolate smoothly between the two limiting
cases.

Both questions can be studied non-perturbatively on the lattice. In order
to have a sensible plane-plane correlation function,
we define the operator~\cite{phtn}
\be
O_{\bf p}(z)=\sum_{x_1,x_2} 
	[\mathop{\rm Im} u_{12}(x_1,x_2,z)]\exp(i {\bf p}\cdot{\bf x}),
\label{def}
\ee
where the sum is taken over the plane $(x_1,x_2)$, $u_{ij}$ is the
plaquette corresponding to the abelian U(1)-field, and ${\bf p}$ is a
transverse momentum vector restricted to plane $(1,2)$: $(p_1,p_2,p_3)
= 2 \pi/N (n_1,n_2,0)$ with integer $n_i$ (there is quantization of
the momentum because of the periodic boundary conditions). In the
continuum limit the operator $O$ is just
\be
O_{\bf p}(z) = \frac{1}{\sqrt{a \beta_G'}}
\int dx_1 dx_2 B_{12}(x_1,x_2,z)\exp(i {\bf p}\cdot{\bf x})
\label{contdef}
\ee
Note that if ${\bf p} = 0$, \eq\nr{contdef} becomes identically zero in the absence of the winding modes.

The correlator of two plane operators,
\be
G(z) = \frac{1}{N^3}\langle \sum_t O_{\bf p}(t)O^*_{\bf p}(z+t)\rangle
\ee
has the long-distance behaviour
\be
G(z) = \frac{A_{\gamma}}{2\beta_G'}\frac{a p^2}{\sqrt{p^2+m_{\gamma}^2}}
\exp(-z \sqrt{p^2+m_{\gamma}^2}),
\la{photoncor}
\ee
where $m_{\gamma}$ is the photon mass and $A_{\gamma}$ is measuring
the projection of the lowest mass state to the hypercharge U(1)
field.  
In the tree approximation we have $A_{\gamma}= 1$ in the
``symmetric'' phase and $A_{\gamma}= \cos^2\!\theta_W$ in the ``broken''
phase. 
For simplicity, \eq\nr{photoncor} is written in terms of the
continuum dispersion relation instead of the lattice one; 
due to the small values of {\bf p} used here the effect the effect
of the different dispersion relations is well below the statistical
errors in this analysis.
As one can see, the use of a momentum dependent operator in (\ref{def})
is essential for a non-zero result for the correlation function.

\begin{figure}[tb]
\figtopspace \vspace{1cm}
\epsfysize=\figysize
\centerline{\epsffile{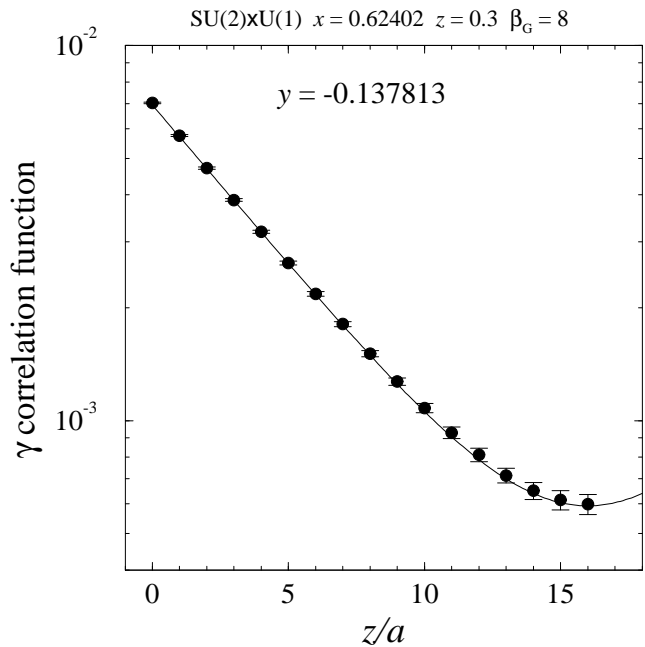}}
\figbottomspace
\caption[a]{The photon correlation function \eq\nr{photoncor} for
$x=0.62402$ at the cross-over value for $y$, for a lattice of size
$32^3$.  The continuous line is a 2-parameter fit to the distance
range 4--16.\la{fig:gammacorr}}
\end{figure}

\begin{figure}[tb]
\figtopspace \vspace{1cm}
\epsfysize=\figysize
\centerline{\epsffile{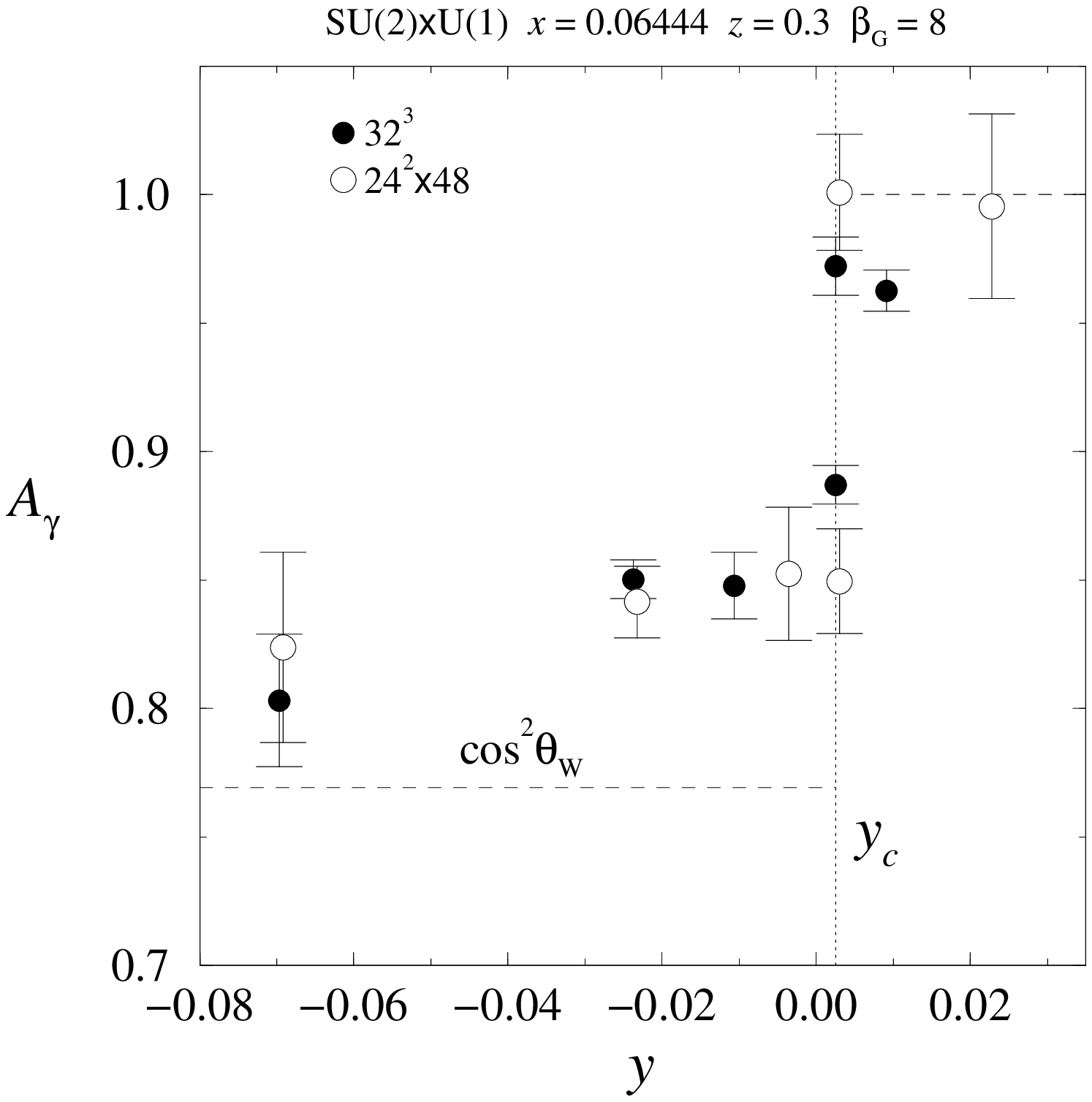}}
\figbottomspace
\caption[a]{The pre-exponential factor $A_\gamma$ for the $x=0.06444$
case.  The dashed horizontal lines are the tree-level values for
$A_\gamma$ in the symmetric and broken phases.\la{fig:preexp}}
\end{figure}

In the simulations we used the two smallest possible transverse
momenta, ${\bf p} = 2\pi/N\,{\bf e_i}$, with $i=1$ or 2.  In
\fig\ref{fig:gammacorr} we show the photon correlation function,
measured from a $x=0.62402$, $\beta_G=8$ -system; all other parameter
values yield very similar correlation functions.  The results are perfectly
consistent with the existence of the massless vector state.  Applying
\eq\nr{photoncor}, the individual fits yield randomly
positive or negative values for $m_\gamma^2$ (compatible with zero within
the statistical errors); as a final result we give an upper bound for the
photon mass over the whole parameter range:
\be
  m_\gamma \lsim 0.03 g_3^2.
\ee

As to the mixing parameter $A_\gamma$, 
in the vicinity of the phase transition with $x=0.06444$ we
observe a sharp discontinuity in
accordance with expectations (see \fig\ref{fig:preexp}).  Note that
here the discontinuity in $A_\gamma$ is considerably smaller than
given by the tree-level formula. This can to some extent
be understood perturbatively; indeed, it is relatively
easy to calculate at 1-loop order the residy 
of the $1/k^2$ pole in the U(1) correlator $\langle B_i B_j\rangle$
(this determines $A_\gamma$ through~\nr{contdef}--\nr{photoncor}).
In the symmetric phase ($m_3^2 > 0$), one gets
\be
A^{\rm symm}_\gamma = 1-\frac{g_3'^2}{48 \pi m_3}=
1-\frac{z}{48 \pi \sqrt{y}},
\ee
whereas in the broken phase the result is 
\be
A^{\rm broken}_\gamma=
\cos^2\!\theta_W
\biggl(1+\frac{11}{12}\frac{e_3^2}{\pi m_W}
\biggr), 
\ee
where $e_3^2=g_3^2 \sin^2\!\theta_W=g_3^2g_3'^2/(g_3^2+g_3'^2)$
and $m_W$ is the perturbative W mass. 
It can be verified that these expressions are gauge-independent.
Both corrections have the right sign and
even roughly the right magnitude. 
In the symmetric phase, the expansion breaks down 
close to the transition point as $y\sim 0$.

As one approaches the endpoint of the line of the first order transitions, 
the discontinuity of $A_\gamma$ gets smaller (together with $m_W$).
When $x=0.62402$ (which is safely above the endpoint) we observe a
continuous change in the mixing parameter $A_\gamma\sim 1$
across the cross-over region.
Thus, our non-perturbative analysis confirms within 
statistical errors the existence of the
massless state which follows from perturbation theory, and
demonstrates the absence of jumps of the mixing parameter in the
region of large Higgs masses.

\subsection{Absence of spontaneous parity breaking}

A priori, it is not excluded that the Euclidean 3d non-abelian
gauge-Higgs system exhibit spontaneous parity breaking in the
strongly coupled phase. Arguments in favour of this possibility,
based on the computation in \cite{topological} of the renormalization
of the 3d topological mass term, were given in \cite{versus}. 
Lattice simulations in the symmetric phase of the SU(2)+Higgs
system were carried out in \cite{parlatt}, where no signal for
spontaneous parity breaking was found. Here we extend the analysis of
\cite{parlatt} to the more realistic SU(2)$\times$U(1) case.

Consider some local gauge invariant operator $O$ which is odd under
the parity transformation. If there is spontaneous parity breaking in
the system, the probability distribution of the quantity
\be
\frac{1}{V}\int d^3 x\, O(x),
\ee
where the integral is taken over a finite volume $V$,
must have a double-peak structure, which becomes more pronounced
when the volume of the system increases. Simultaneously, the
susceptibility defined as
\begin{equation}
\chi(V,T)=\int d^3x\langle {{O}}(x){{O}}(0)\rangle,
\end{equation}
must behave as $\chi(V, T_c) \propto V^1$ when $V\rightarrow \infty$.
If spontaneous parity breaking is absent, the probability
distribution of $O$ looks like a single peak with center at zero, and
$\chi(V, T_c) \propto V^0$.

Our choice of parity breaking operators includes the the lattice
versions of the operators (\ref{parityop1}--\ref{parityop2})
and the operator
\be
  W_P(x) = \epsilon_{ijk} W^3_i(x) W^+_j(x) W^-_k(x),
\ee
where $W^\pm \equiv W^1 \pm iW^2$. The study of the non-local
operators (\ref{csop}) corresponding to the Chern-Simons numbers, 
was not attempted because it would be very time consuming.

We found that the expectation values of all of these operators remain
consistent with zero in both symmetric and broken phases at $x=
0.06444$ near the phase transition.  More precisely, the distribution
of the operators is very accurately Gaussian centered around zero;
moreover, the autocorrelations of the successive measurements are
negligible even in the transition region, where all of the other
measurements show significant autocorrelations.  This implies that the
parity operators decouple from the long-distance dynamics and are
sensitive only to the local ultraviolet noise of the system.  The
susceptibility $\chi$ was found to be practically volume independent
for all of the volumes, again signifying the non-critical nature of
the parity operators.  The situation is similar at $x=0.62402$ in the
cross-over region.  Thus, the possibility of non-perturbative parity
breaking is ruled out (at least within the statistical accuracy and
for the volumes used in our simulations) for the
SU(2)$\times$U(1) theory as well.

\begin{figure}[tb]
\figtopspace
\centerline{\epsfysize=\figysize \epsfbox{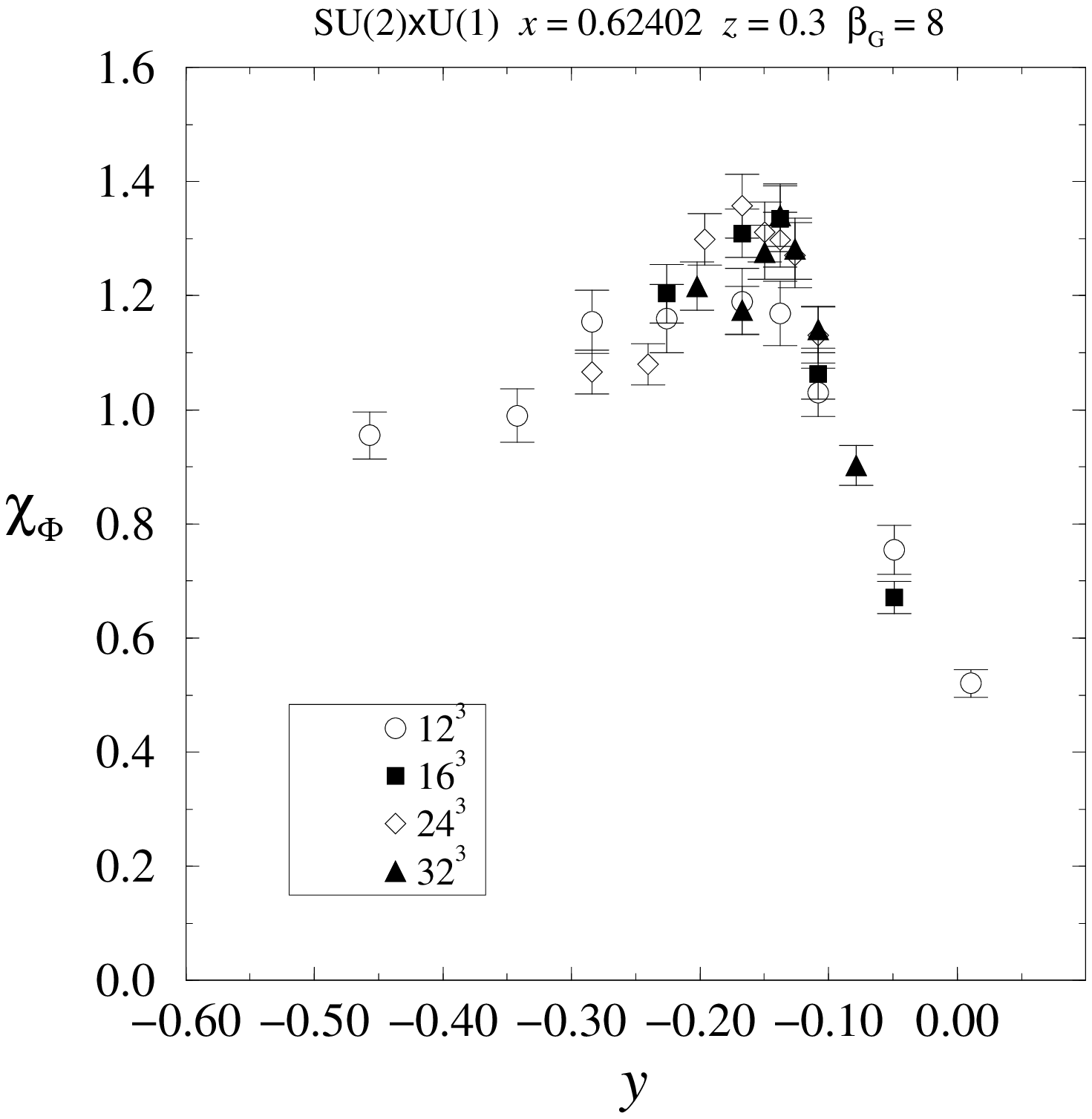}}
\vspace*{-6cm}
\centerline{\epsfysize=\figysize \epsfbox{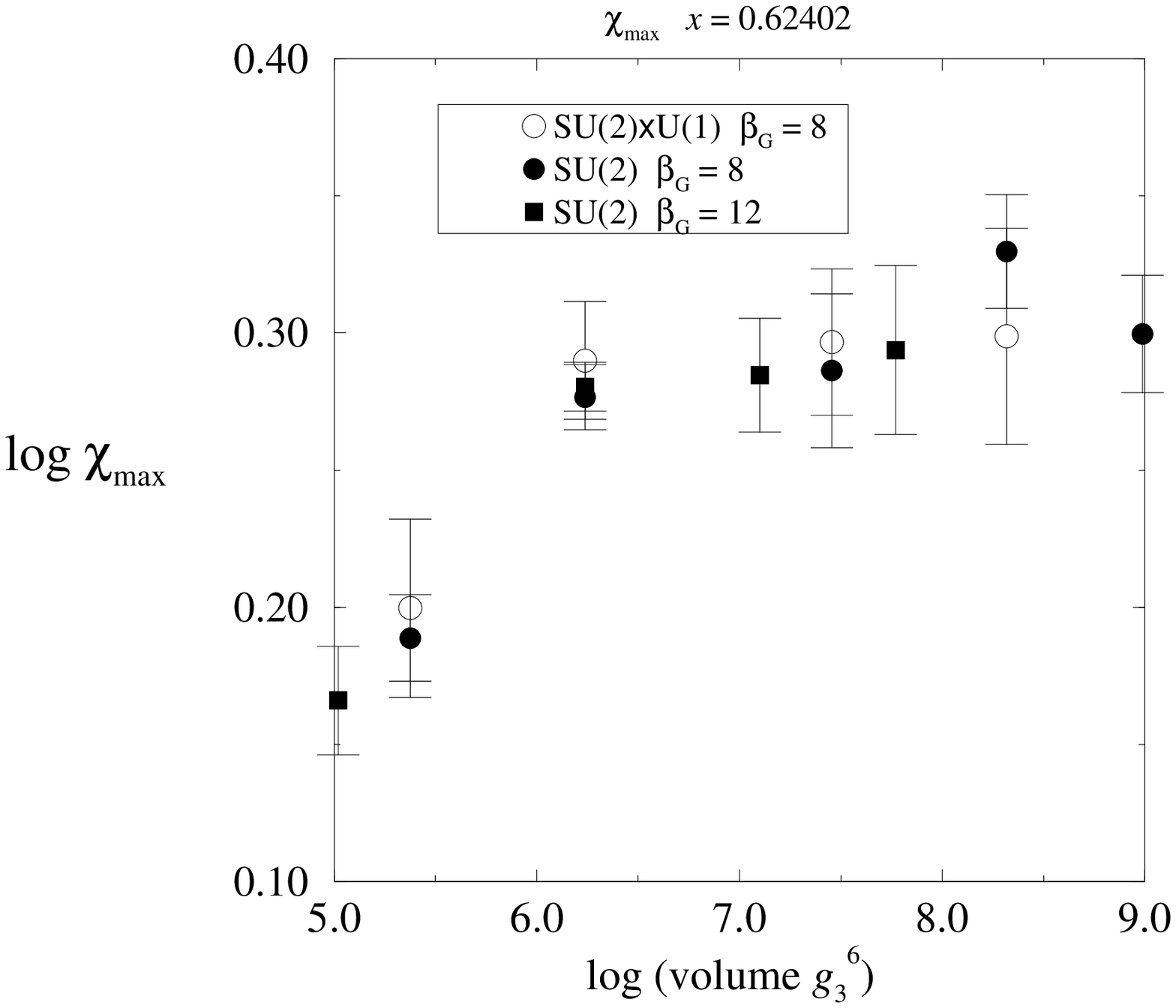}}
\figbottomspace
\caption[a]{The $\phi^\dagger\phi$ susceptibility $\chi_\phi$
measured
from $\beta_G=8$, $x=0.62402$ lattices as a function of $y$.  No
divergence develops when the volume increases (top). 
The maximum values of the susceptibility $\chi_\phi$
as a function of the volume for SU(2)+Higgs and
SU(2)$\times$U(1)+Higgs
systems (bottom).\la{m180chimax}
}
\end{figure}

\subsection{The transition in the $x=0.62402$ case}

When $x=0.62402$ (in
\cite{isthere?} this value of $x$ corresponds to $m_H^*=180$\,GeV)
and $z=0.3$
the transition has turned into a smooth and
regular cross-over, completely analogously to the SU(2)+Higgs 
$z=0$ case.
This can be observed, for example, by monitoring the behaviour of
the order parameter susceptibility and the correlation lengths.
For example, in \fig\ref{m180chimax} we show the behaviour of the
$\phi^\dagger\phi$ susceptibility $\chi_\phi$ as a function of
$y$ for several volumes; in the case of a true transition the
susceptibility diverges at the transition point ($\propto V$ for a
first order transition, $\propto V^\gamma$ for a second order one).
In this case no divergence is seen, only a volume independent
smooth peak corresponding to the cross-over behaviour.  For this
$x$ we have data only for $\beta_G=8$, and using the maximum location
of $\chi_\phi$ we extrapolate the infinite volume
cross-over coupling to be $y_c(x=0.62402,z=0.3,\beta_G=8) = -0.152(7)$.

In \fig\ref{m180chimax} we compare the maximum values of
$\chi_\phi$ for SU(2)+Higgs and SU(2)$\times$U(1)+ 
Higgs theories at $x=0.62402$; the SU(2)+Higgs data is from \cite{isthere?}.
The quantitative behaviour of the systems is quite similar, 
and there is no first order transition.
In \fig\ref{um180mh} we show the (lightest) scalar mass across the 
transition region; it remains non-zero as in the SU(2)+Higgs, 
signalling the absence of a second order transition.
The photon mass vanishes everywhere 
in both phases (see Sec.~\ref{photon}), and hence
it should not affect the order of the transition.

\begin{figure}[tb]
\figtopspace \vspace{1cm}
\epsfysize=\figysize
\centerline{\epsffile{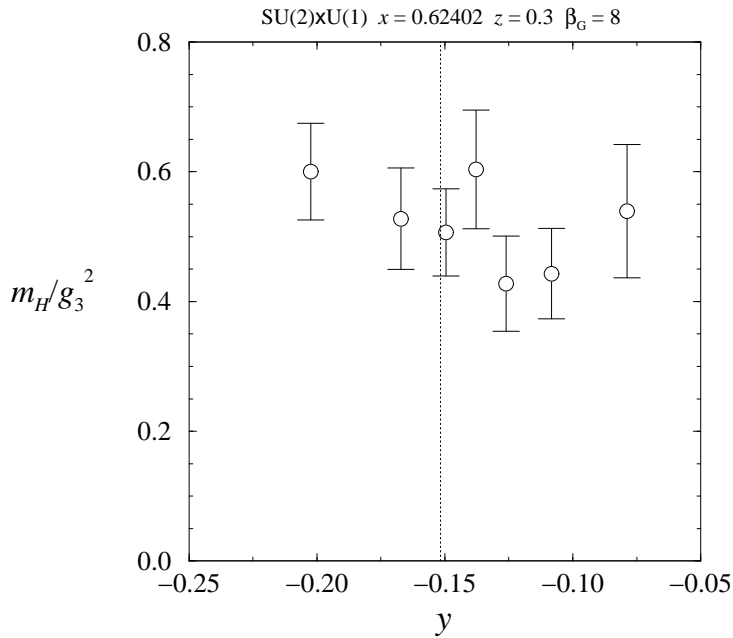}}
\figbottomspace
\caption[a]{The scalar correlation length
from a $\beta_G=8$, $x=0.62402$, volume $=32^3$ 
lattice as a function of $y$.
The dotted line shows the location of the maximum of $\chi_\phi$.
\la{um180mh}}
\end{figure}

\section{Conclusions}

We have in this paper studied numerically the
3d SU(2)$\times$U(1)+fundamental Higgs theory and compared it
with the 3d SU(2)+fundamental Higgs theory at the same values of the 
coupling $x=\lambda_3/g_3^2$, but for
$g_3'^2\not=0$. 
The comparison has been made both in a region with a
clear first-order transition ($x=0.06444; m_H^*=60$ GeV)
and in the cross-over region ($x=0.62402; m_H^*=180$ GeV). The
main conclusion is that the inclusion of U(1) does not bring
in significant new non-perturbative effects: the changes in the
non-perturbatively computed quantities due to increasing
$g_3'^2$ are as large as those given by perturbation theory.
At the same time, the values of the observables themselves, 
especially $\sigma_3$, differ from those in perturbation theory.
In particular, the first order transition ends at some 
$x_c$ as in 3d SU(2)$+$Higgs and for $x>x_c$ there is 
no first or second order
transition. A still higher order transition cannot
be strictly excluded by the analysis presented
in this paper.

We have also confirmed non-perturbatively the existence of a
(nearly) massless photon in the 3d SU(2)$\times$U(1)+Higgs model.
We have studied  
the mixing of this state to the Lagrangian U(1)-field,  
and demonstrated that the jump in the mixing parameter
vanishes together with the first order transition 
as one goes to large enough Higgs masses. Physically these
results on the photon imply that magnetic fields are not screened in hot
electroweak plasma even taking into account all nonperturbative effects.

\section*{Acknowledgements}

M.L was partially supported by the University of Helsinki.


\end{document}